%% file: ms.tex
\documentclass[12pt,preprint]{emulateapj}

\shorttitle{The temperatures of Red Supergiants}
\shortauthors{Davies et al.}
\usepackage{graphicx}
\usepackage{natbib}
\usepackage{lscape,array}
\def\ga{\mathrel{\hbox{\rlap{\hbox{\lower4pt\hbox{$\sim$}}}\hbox{$>$}}}}
\def\la{\mathrel{\hbox{\rlap{\hbox{\lower4pt\hbox{$\sim$}}}\hbox{$<$}}}}

\def\msun{$M$\mbox{$_{\normalsize\odot}$}}

\def\lsun{$L$\mbox{$_{\normalsize\odot}$}}
\def\lbol{$L$\mbox{$_{\rm bol}$}}
\def\kms{\,km~s$^{-1}$}

\def\arcsec{$^{\prime \prime}$}
\def\arcmin{$^{\prime}$}
\def\teff{$T_{\rm eff}$}
\def\Teff{$T_{\rm eff}$}
\def\logg{$\log g$}

\def\um{$\mu$m}
\def\chisq{$\chi^{2}$}
\def\AV{$A_{V}$}
\def\hminus{H$^{-}$}
\def\Hminus{H$^{-}$}
\def\ebmv{$E(B-V)$}

\newcommand{\fig}[1]{Fig.\ \ref{#1}}

\input{journals}

\begin{document}
\title{The temperatures of Red Supergiants} 
  \author{Ben Davies\altaffilmark{1,2,3},
  Rolf-Peter Kudritzki\altaffilmark{4}, Bertrand Plez\altaffilmark{5},
  Scott Trager\altaffilmark{6}, Ariane Lan{\c c}on\altaffilmark{7}, \\Zach
  Gazak\altaffilmark{4}, Maria Bergemann\altaffilmark{8}, Chris
  Evans\altaffilmark{9}, Andrea Chiavassa\altaffilmark{10} }

\affil{$^{1}$Astrophysics Research Institute, Liverpool John Moores 
University, Egerton Wharf, Birkenhead, CH41 1LD, UK.}

\affil{$^{2}$Institute of Astronomy, University of Cambridge,
  Madingley Road, Cambridge CB3 0HA, UK.}

\affil{$^{3}$School of Physics \& Astronomy, University of Leeds,
  Woodhouse Lane, Leeds LS2 9JT, UK.}

\affil{$^{4}$Institute for Astronomy, University of Hawaii, 2680
Woodlawn Drive, Honolulu, HI, 96822, USA} 

\affil{$^{5}$Laboratoire Univers et Particules de Montpellier,
  Universit\'{e} Montpellier 2, CNRS, F-34095 Montpellier, France}

\affil{$^{6}$Kapteyn Institute, University of Groningen, P.O. Box 800,
  9700 AV Groningen, The Netherlands}
\affil{$^{7}$Observatoire astronomique and CNRS UMR 7550, Université
  de Strasbourg, Strasbourg, France}
\affil{$^{8}$Max-Planck-Institute for Astrophysics,
  Karl-Schwarzschild-Str.1, D-85741 Garching, Germany}
\affil{$^{9}$UK Astronomy Technology Centre, Royal Observatory
  Edinburgh, Blackford Hill, Edinburgh., EH9 3HJ, UK; Institute for
  Astronomy, Royal Observatory Edinburgh, Blackford Hill, Edinburgh.,
  EH9 3HJ, UK}
\affil{$^{10}$Universit\'{e} de Nice Sophia-Antipolis, Observatoire de la C\^{o}te d'Azur, CNRS Laboratoire Lagrange, BP 4229, 06304, Nice Cedex 4, 
France}

\begin{abstract}
We present a re-appraisal of the temperatures of Red Supergiants
(RSGs) using their optical and near-infrared spectral energy
distributions (SEDs). We have obtained data of a sample of RSGs in the
Magellanic Clouds using VLT+XSHOOTER, and we fit MARCS model
atmospheres to different regions of the spectra, deriving effective
temperatures for each star from (a) the TiO bands, (b) line-free
continuum regions of the spectral energy distributions (SEDs), and (c)
the integrated fluxes. We show that the temperatures derived from fits
to the TiO bands are systematically {\it lower} than the other two
methods by several hundred Kelvin.  The TiO fits also dramatically
over-predict the flux in the near-IR, and imply extinctions which are
anomalously low compared to neighbouring stars. In contrast, the SED
temperatures provide good fits to the fluxes at all wavelengths other
than the TiO bands, are in agreement with the temperatures from the
flux integration method, and imply extinctions consistent with nearby
stars. After considering a number of ways to reconcile this
discrepancy, we conclude that 3-D effects (i.e.\ granulation) are the
most likely cause, as they affect the temperature structure in the
upper layers where the TiO lines form. The continuum, however, which
forms at much deeper layers, is apparently more robust to such
effects. We therefore conclude that RSG temperatures are much warmer
than previously thought. We discuss the implications of this result
for stellar evolution and supernova progenitors, and provide
relations to determine the bolometric luminosities of RSGs from
single-band photometry.

\end{abstract}



\section{Introduction} \label{sec:intro}

Red Supergiants (RSGs) are a post main-sequence (MS) phase of stars
with masses
$\sim$8-30\msun\ \citep[e.g.][]{Mey-Mae00,Eldridge08,Brott11}. Their
luminosities \citep[$\ga10^{4.5-5.8}$\lsun][]{H-D79} rival those of
globular clusters and dwarf galaxies, and dominate the light output of
their host galaxies at near-infrared (IR) wavelengths. For most stars
in this mass range, the RSG phase is thought to be the evolutionary
stage immediately preceeding core-collapse supernova (SN), though
searches of pre-explosion images are yet to find a RSG progenitor with
an inferred initial mass greater than
$\sim$20\msun\ \citep[][]{Smartt09,Fraser12}.

In determining the bolometric luminosities of RSGs, knowledge of the
temperature scale is crucial, since for the typical temperature range
of RSGs (\Teff$\sim$3400-4500K) the optical bolometric correction can
vary by almost 2 magnitudes \citep{Levesque05}. Therefore, in order to
accurately convert the flux in a given photometric pass-band to the
total flux, we must know the relationship between the star's observed
properties (e.g. photometric colours, spectral type) and its
\Teff. This problem is complicated further by the fact that extinction
\AV, whether interstellar or circumstellar, can make the star appear
cooler, introducing a large degree of degeneracy between \AV\ and
\Teff.


To measure the effective temperatures of RSGs, a number of different
strategies have been employed, though all methods ultimately require a knowledge of the radiative transfer through the star's atmosphere. One method is to measure the angular diameters of stars through interferometry or Lunar occultations \citep{Lee70,Dyck96,Mozurkewich03,vanBelle09}. Since the definition of \Teff\ comes from $L_{\star}=4 \pi R_{\star}^2 \sigma T_{\rm eff}^4$, where $L_{\star}$ and $R_{\star}$ are the stellar luminosity and radius respectively, by measuring the apparent sizes and fluxes one can determine \Teff. 

This method does have a number of problems however. Firstly, the measured size of the star is very sensitive to the observed wavelength, with the star appearing larger in spectral regions of high opacity. For example, interferometric
observations in the near-IR have shown that the apparent size of the
star may increase by up to 50\% around the wavelengths of strong
molecular lines compared with the nearby continuum
\citep{Perrin04,Ohnaka09,Tsuji00,Ohnaka11,Chiavassa10}. Therefore, in order to interpret the measurements of angular diameters, model stellar atmosphere are required. Early work typically used black-bodies, which are well-known to be a poor representation of RSGs, particularly in the optical. Later work has typically used plane-parallel Kurucz models, again known to be inadequate for RSGs. 

Secondly, since one needs to know the flux of the star, one also needs to know the foreground extinction. Typically it is assumed that the target star has the same foreground extinction as the other stars in its host cluster or association. However, this is likely to be an underestimate, since RSGs can produce their own circumstellar dust through their winds, providing up to a magnitude of extra visual extinction \citep[e.g.][]{Schuster06,deWit08}. A combination of these two effects could explain why near-IR angular diameter measurements provide warmer effective temperatures (i.e.\ smaller stellar radii) than those in the optical \citep{vanBelle09}.

Other investigations into the temperatures of RSGs have looked at their spectral energy distributions (SEDs). \citet{Dyck74} defined a near-IR `colour temperature', work which was built upon by \citet{Flower75,Flower77}. This work was however calibrated this against angular diameter measurements taken at 7100\AA, which is at the centre of a deep absorption feature, and so may have over-estimated the radii. Later work fit data with synthesised spectra from plane-parallel models \citep{W-W78,S-S79}, though these models lacked detailed molecular opacities. 

The most contemporary measurement of the RSG temperature scale is presented in \citet[][ hereafter L05 and L06]{Levesque05,Levesque06}. These authors obtained
spectrophotometry of samples of RSGs in the Galaxy and the Magellanic
Clouds in the {\it BVRI} region of the spectrum. They then fitted this
region of the spectrum, which is dominated by the TiO absorption bands
which determine the spectral classification of these stars, with MARCS
model atmospheres \citep{Gustafsson08}, which are 1-D spherical hydrostatic
models, important for stars with extended atmospheres. The temperature
scale they found for Galactic stars was slightly warmer than was
previously thought. These warmer temperatures brought Galactic RSGs
into closer agreement with the predicted Hayashi limit of Geneva
stellar models\footnote{It should be noted that the location of the
  Hayashi limit in models of massive stars is governed by the
  convective mixing length parameter, which is typically assumed to be
  the same as for the Sun. Observations of globular clusters have
  shown that a uniform mixing length may not be appropriate for all
  stars, especially cool stars \citep{Ferraro06}.}  \citep{Mey-Mae00}.

There are, however, known issues and anomalies with the L05 \Teff\ scale as
well. These authors found that the temperatures derived from fitting
the optical spectra were systematically offset from those expected
from the $V-K$ colours, indicating that their model fits were not
reproducing the correct near-IR flux. They also found a number of very
cool stars in the SMC, whose temperatures were so low that they were
apparently in violation of hydrostatic equilibrium
\citep{Levesque07,Massey07}. In addition, \citet{Lancon07} fit the
optical and near-IR spectra of a sample of Galactic RSGs with PHOENIX
models, finding problems simultaneously fitting the strengths of
various molecular bands, particularly for the most luminous stars.

The reasons for these discrepancies could lie in the detailed
radiative transfer that governs the strengths of the TiO bands, which
are used as temperature indicators by L05 and L06. As will be
discussed later, the strengths of these bands are sensitive to a
number of factors, such as the temperature structure of the atmosphere
and metallicity. Such factors may make the TiO bands unreliable
temperature indicators.
 
Here we explore an alternative method of deriving the temperatures of
RSGs. As in L05 and L06, we fit spectra with MARCS model
atmospheres. However, in addition to looking at the strengths of the
TiO bands, we also investigate the \Teff\ implied by fits to the
optical-infrared SED using regions which are free of the deep
molecular absorption seen in the optical regions. We do this on a
sample of stars taken from the two Magellanic Clouds, which should
have relatively low foreground extinction, and where the discrepancies
in the Levesque et al. temperature scale appear to be most conspicuous
\citep{Levesque07}.

We begin in Sect.\ \ref{sec:obs} with a description of our
observations and data reduction. In Sect.\ \ref{sec:anal} we discuss
our analysis techniques, and present our results in
Sect.~\ref{sec:results} which indicate a systematic error in the L06
temperature scale. Reasons for this discrepancy, and possible
remedies, are discussed in Sect.\ \ref{sec:disc}, as well as the
implications of our results. We conclude in Sect.\ \ref{sec:conc}.

\begin{table*}
  \caption{Observation data for the stars in our sample. The errors on
    the synthetic photometry are $\pm$0.1mag for the optical data and
    $\pm 0.04$mag for the infrared data.}
  \centering
  \begin{tabular}{ccccccccccccc}
\hline \hline
Star  & RA DEC & Obs date & $t_{\rm UVB}$ & $t_{\rm VIS}$ & $t_{\rm NIR}$ & 
$B$ & $V$ & $R$ & $I$ & $J$ & $H$ & $K_{S}$ \\
  & (J2000) &   & (sec) & (sec) & (sec) &  &  &  &  &  &  &  \\ 
\hline
SMC 011709 & 0 48 46.32  -73 28 20.7 & 2011-10-13 & 60 & 10 & 50 & 14.43 & 
12.60 & 11.63 & 10.76 & 9.63 & 8.84 & 8.61 \\
SMC 013740 & 0 49 30.34  -73 26 49.9 & 2011-12-06 & 80 & 12 & 70 & 15.56 & 
13.78 & 12.76 & 11.86 & 10.61 & 9.75 & 9.49 \\
SMC 020133 & 0 51 29.68  -73 10 44.3 & 2011-12-06 & 60 & 10 & 50 & 14.84 & 
12.86 & 11.76 & 10.73 & 9.39 & 8.56 & 8.27 \\
SMC 021362 & 0 51 50.25  -72 05 57.2 & 2011-12-05 & 80 & 12 & 70 & 14.91 & 
13.02 & 12.00 & 11.09 & 9.91 & 9.05 & 8.80 \\
SMC 030616 & 0 54 35.90  -72 34 14.3 & 2011-12-06 & 60 & 10 & 50 & 14.50 & 
12.67 & 11.65 & 10.69 & 9.48 & 8.68 & 8.45 \\
SMC 034158 & 0 55 36.58  -72 36 23.6 & 2011-12-06 & 80 & 12 & 70 & 14.87 & 
13.01 & 12.01 & 11.14 & 10.04 & 9.26 & 9.03 \\
SMC 035445 & 0 55 58.84  -73 20 41.4 & 2011-12-06 & 100 & 15 & 100 & 14.59 & 
12.91 & 12.00 & 11.20 & 10.09 & 9.31 & 9.09 \\
SMC 049478 & 1 00 41.56  -72 10 37.0 & 2011-12-03 & 50 & 5 & 30 & 14.23 & 
12.40 & 11.35 & 10.36 & 9.20 & 8.42 & 8.15 \\
SMC 050840 & 1 01 15.99  -72 13 10.0 & 2011-12-06 & 60 & 10 & 50 & 14.62 & 
12.72 & 11.66 & 10.69 & 9.48 & 8.64 & 8.38 \\
SMC 057386 & 1 03 47.35  -72 01 16.0 & 2011-12-30 & 80 & 12 & 70 & 14.22 & 
12.71 & 11.83 & 11.07 & 10.01 & 9.24 & 9.01 \\
LMC 064048 & 5 04 41.79  -70 42 37.2 & 2011-12-05 & 60 & 10 & 50 & 15.09 & 
13.15 & 11.94 & 10.69 & 9.42 & 8.55 & 8.19 \\
LMC 067982 & 5 05 56.61  -70 35 24.0 & 2011-12-05 & 60 & 10 & 50 & 14.80 & 
12.84 & 11.73 & 10.61 & 9.27 & 8.40 & 8.11 \\
LMC 116895 & 5 19 53.34  -69 27 33.4 & 2011-12-06 & 60 & 10 & 50 & 14.49 & 
12.62 & 11.54 & 10.44 & 9.17 & 8.39 & 8.08 \\
LMC 131735 & 5 23 34.09  -69 19 07.0 & 2011-11-29 & 80 & 12 & 70 & 14.33 & 
12.60 & 11.73 & 10.98 & 9.87 & 9.15 & 8.95 \\
LMC 137818 & 5 27 14.33  -69 11 10.7 & 2011-11-29 & 60 & 10 & 50 & 15.34 & 
13.38 & 12.03 & 10.65 & 9.34 & 8.52 & 8.18 \\
LMC 142202 & 5 28 45.59  -68 58 02.3 & 2011-12-06 & 50 & 5 & 30 & 14.25 & 
12.30 & 11.25 & 10.19 & 8.78 & 7.95 & 7.64 \\
LMC 143877 & 5 29 21.10  -68 47 31.5 & 2011-10-13 & 50 & 5 & 30 & 14.04 & 
12.13 & 11.13 & 10.21 & 9.00 & 8.28 & 8.01 \\
LMC 158317 & 5 33 44.60  -67 24 16.9 & 2011-10-13 & 80 & 12 & 70 & 15.12 & 
13.13 & 12.01 & 10.97 & 9.64 & 8.81 & 8.48 \\
\hline
  \end{tabular}
  \label{tab:obs}
\end{table*}

\section{Observations \& data reduction} \label{sec:obs}
We have obtained observations of several stars in the LMC and SMC
using VLT+XSHOOTER \citep{XSHOOTER} under ESO programme number
088.B-0014(A) (PI B.\ Davies). The sample of stars, 8 in the LMC and 
10 in the SMC, were selected from L06. We aimed to sample the full 
distribution of spectral types observed in each galaxy. We also deliberately 
excluded stars which may be `extreme' objects, such as analogues of the 
Galactic RSG VY~CMa. We did so by avoiding stars which had luminosities in 
excess of $10^{5.5}$\lsun\ in L06.

The stars were observed in nodding ABBA
mode with four exposures per star, and a randomized jitter at each
position on the slit. For each of the instrument arms -- UVB, VIS, and
NIR -- we used the 5.0\arcsec slits to minimize slit losses and obtain
accurate spectrophotometry. The spectral resolution was therefore set
by the seeing, which for seeing of 1\arcsec\ was roughly $R \equiv
\lambda / \Delta \lambda \sim 5000$ at all wavelengths. The precise
value of $R$ was determined at the analysis stage (see
Sect. \ref{sec:anal}). Integration times are listed in Table
\ref{tab:obs}, and were chosen to achieve a signal-to-noise ratio
(SNR) of at least 50 per resolution element at all wavelengths greater
than 400nm. Since the UVB and VIS arms take much longer to read-out
than the NIR arm, and each arm operates independently, extra time
could be spent integrating in the NIR.

Flux standard stars were observed each night and to correct for the
atmospheric absorption in the NIR, telluric standard stars of spectral
type late-B were observed within one hour of each science target. In
general observing conditions were good on each night, and the seeing
was always $<$1\arcsec. The standard suite of XSHOOTER calibration
frames used in the data reduction process were taken at the beginning
and end of the night \citep[for details, see][]{Goldoni06}.

The initial steps of the reduction process were done using the
XSHOOTER data reduction pipeline \citep{Modigliani10}. These steps
included subtraction of bias and dark frames, flat-fielding, order
extraction and rectification, flux and wavelength calibration. The
accuracy of the wavelength solution was checked by measuring the
residuals of the wavelengths of the arc lines. The root-mean-square of
these residuals was found to be below $\sim$0.1 pixels ($\sim$1\kms)
for all targets. To check the flux calibration we applied the response
function to the flux standard and compared with the calibration
spectrum. We found residuals of order 1\%. The absolute accuracy of
the flux calibration is discussed in Sect.\ \ref{sec:synthphot}. When
stitching the spectra of the three different arms together, we clipped
the UVB arm at 563nm (to avoid the flux dropout due to the first
dichroic), the VIS arm between 558-1010nm, and the NIR arm below
2350nm. The overlapping regions had consistent fluxes to within 1\%,
and the flux at these regions was calculated as the mean of the two
overlapping spectra. 

The spectra of the science targets and the telluric standards were
then extracted from final rectified two-dimensional orders. The NIR
spectra of the telluric standards were found to have some non-gaussian
noise, mainly from bad pixels and cosmic-ray hits. To identify and
remove this noise a new spectral extraction algorithm was
written. Briefly, bad pixels were flagged by comparing the spatial
profile at a given spectral channel with the median spatial profile of
the neighbouring 3 spectral channels either side. These pixels were
then replaced with the corresponding value in the median
profile. Typically, around five such pixels were found and replaced
per spectral order.

\subsection{Telluric correction}
The method of removing the telluric absorption features depended on
the observing arm. For the UVB arm, telluric features are only present
at the blue end of the wavelength range, where the SNR of our science
data was very low due to their late spectral types. For this reason,
no telluric correction was performed for the UVB arm. 

In the VIS arm, telluric features are non-negligible, with prominent
telluric features at 686nm, 717nm, 759nm, 815nm, and 900nm. To correct
the spectra in each of these regions, we used a synthetic telluric
spectrum, rather than the standard stars which have many intrinsic
spectral features at these wavelengths, such as the Balmer hydrogen
lines. The synthetic spectrum was optimized to best match the telluric
absorption in an iterative process. In each region the relative shift
of the synthetic spectrum with respect to the science data was found
by cross correlation. The strength of the absorption features and
effective resolution was then tuned to give the best
cancellation. This whole process was then repeated.


In the NIR arm, a similar process was employed, only this time the
spectrum of the telluric standard star was used. As we wish to retain
flux calibration of the science data, we first flattened the telluric
spectrum to remove the intrinsic continuum slope, and removed the
Brackett hydrogen lines by fitting voigt profiles. The stellar
continuum was corrected by fitting a black-body curve to the regions
free of telluric absorption (the same regions used by the XSHOOTER
pipeline in the flux-calibration step of the reduction). The
temperature of the model black-body was chosen to ensure
that the ratio of its spectrum to the flux-calibrated telluric
standard star was closest to unity in all spectral regions tested. The
accuracy of this fit was better than 5\% in all tested regions of the
spectrum.

The remaining spectrum, which was a pure measure of the telluric
absorption, was then tuned to match the science data. To do this, the
data was first split into chunks of width 10nm. For each chunk, the
optimal shift, absorption strength and resolution was found as in the
VIS arm. We then assumed that the absorption strength would be uniform
across the whole spectrum, while the shift and resolution would vary
linearly as a function of wavelength. The absorption strength at all
wavelengths was taken to be the median of all those measured in each
of the spectral chunks. The optimized resolution and wavelength shift
as a function of wavelength were obtained from linear fits to the
results from the separate spectral chunks.

\subsection{Synthetic photometry} \label{sec:synthphot}
To check the accuracy of our photometric calibration, and for
comparison with future and past measurements, we derived synthetic
photometry for all our stars. We convolved our XSHOOTER spectra with
Bessell {\it BVRI} filter profiles, as well as with those of the 2MASS
{\it JH$K_{s}$} filters. 

The dominant sources of error in the synthetic photometry are the
accuracy of the spectrophotometric response function, and for bands
which have significant telluric absorption, the quality of the
telluric correction. Tests of the response function, calculated by the
XSHOOTER reduction pipeline, showed that it should be accurate to a
few percent, providing that the observing conditions did not change
between observing the science targets and the flux
standards. Similarly, imperfections in the telluric correction should
only account for a few percent of the total measured flux. 

To check the photometric precision and accuracy empirically, we
compared our synthetic photometry with that of the Magellanic Clouds
Photometric Survey \citep{Zaritsky02,Zaritsky04} and the 2MASS
point-source catalogue \citep{Cutri03}. Though many of the stars in
our sample are known to be variable, the standard deviation of the
differentials between our photometry and that in the literature can
provide an upper limit to the experimental uncertainties, whilst also
giving an indication of any systematics. We find that the optical {\it
  BVI} photometry has 1$\sigma$ standard deviations of $\pm$0.2mags,
while for the 2MASS {\it JH$K_{s}$} photometry it is
$\pm$0.08mags. The systematic offsets are consistent with being
zero. Given the variable nature of RSGs, which can have amplitudes
$\ga$0.5~mag \citep[e.g.][]{Kiss06,Levesque07,Yang12}, a reasonable
conservative estimate of our photometric errors is around half these
values, or $\pm0.1$mags for the optical data and $\pm 0.04$ for the
infrared.

\section{Analysis} \label{sec:anal}
For this study we have computed a whole new grid of model
atmospheres. These atmospheres were generated using the MARCS code
\citep{Gustafsson08}, which operates under the assumptions of LTE,
spherical symmetry and hydrostatic equilibrium. The grid of models is
four-dimensional, computed with a range of metallicities ($\log
(Z/Z_{\odot}) \equiv [Z]$), gravities (\logg), effective temperatures
(\teff) and microturbulent velocities ($\xi$). All model atmospheres
were computed at $\xi$=2\kms, with synthesized spectra computed at
2\kms\ and 5\kms. The relative abundance ratios were taken from
\citet{Grevesse05}. Synthetic spectra were computed from the model
atmospheres using the {\sc turbospectrum} code \citep{A-P98,Plez12},
with a spectral resolution of $R=500,000$ between 250-2500nm. The
chemical composition was scaled from Solar at [Z]=-1.5 and +0.5 in
steps of 0.25dex; \teff\ between 3400 and 4000K in steps of 100K, with
further models at 4200K and 4400K; \logg\ between -0.5 and +1.0 in
steps of 0.5 (in cgs units). All models were computed with an adopted
stellar mass of $M_{\star}$=15\msun. Though RSGs may have masses
between $\sim$8-25\msun, the pressure scale height remains largely
unchanged throughout this mass range \citep[see discussion in
][]{rsg_jband}. Within this grid of models, we logarithmically
interpolated model spectra at metallicities appropriate for the LMC
and SMC, specifically [Z]=-0.65 and -0.4 \citep{Trundle07}, with
models $\pm$0.25dex either side of these to investigate the effect of
variations in metallicity. We also logarithmically interpolated the
temperatures onto a finer grid of spacing 20K.

To derive the temperatures of the stars, we employ three different
techniques, and compare the results of each. Below we describe the
methodolgy of each. 

\subsection{The TiO method}
Firstly, we extract a region of the spectrum around the TiO bands
between 500-800nm. We compare the flux in this spectral window with
those of a subset of models. We initially selected models at the
appropriate metallicity for the star's host galaxy (see previous
section), \logg=0.0, and $\xi$=2\kms. The choice of gravity is
motivated by the typical masses, luminosities and radii of RSGs
\citep[see e.g.][]{Mey-Mae00}. Microturbulence parameters are
typically between 2-4\kms\ \citep{Cunha07}. In practice, varying
metallicity, \logg\ and $\xi$ was found to have little effect on our
results (see below). The parameters we are therefore allowing to vary
are the temperature \teff\ and the extinction \AV, with the latter
being computed from the reddening law of \citet{Gordon03}. Broadly,
\teff\ affects the strengths of the TiO absorption, while \AV\ affects
the overall slope of the spectrum. We find the best fitting model by
computing the \chisq\ summed over all spectral pixels for each value of 
\teff\ and \AV. We then fit for the two parameters simultaneously by finding the 
model with the minimum \chisq\ in \teff-\AV\ space. This method weights all 
TiO bands within the wavelength range equally.

The errors in \Teff\ and \AV\ are measured from the maximal values of
those models which have \chisq\ values within 0.5 of the best fitting
model -- this was found to be the point at which the quality of the
fit becomes noticeably poorer. We also explored the best fitting
models with [Z]=$\pm$0.25dex, with \logg=-0.5 -- +1.0dex, and with
$\xi$=5\kms. If these fits provided values of \Teff\ and \AV\ which
were outside the errors of the initial fits, then the errors on these
values were replaced by the best fits of these other subgrids. In
general though, these differences were small ($<$50\,K).

\subsection{The SED method}
The second method of measuring the temperatures was to apply the same
technique, but to the whole of the SED {\it except} those regions
dominated by molecular absorption, since the strengths of these bands
may be sensitive to the layer at which they form in the upper
atmosphere. We avoid the whole of the {\it BVR} spectral region, the
deep absorption bands of TiO and VO, the several narrow CO bands at
$\sim$1.6\um\ and the band-heads $>$2.3\um, and the CN band at 1.1\um.

Before fitting, spectra were first smoothed by a boxcar filter of
width 10 pixels in order to wash out narrow spectral features, as it
is the broad shape of the SED that we are using as a diagnostic of the
temperature. As with the TiO method, we used a \chisq-minimization
technique to find the best-fitting values of \teff\ and \AV. We again
used the reddening law of \citet{Gordon03}, and explored the effects
of varying gravity, abundance and microturbulence in computing our
error estimates. We also experimented with different extinction 
laws, specifically those of \citet{Cardelli89} and \citet{R-L85}, as well as 
values of $R_V$ between 2-6 for the Cardelli et al.\ law. We found that 
varying the extinction law had only a very minor effect on our derived 
temperatures, resulting in differences to our best-fit temperatures which were less than $\pm$50K. 

\subsection{The Flux Integration method}
 The third method uses the definition of the effective temperature,

\begin{equation}
\sigma T_{\rm eff}^{4} = \pi\int_{0}^{\infty} F_{\lambda}d\lambda
\label{equ:fim}
\end{equation}

\noindent by a wavelength integral over the stellar surface flux,
where $\sigma$ is the Stefan-Boltzmann constant. This method takes
advantage of XSHOOTER's wide wavelength range which allows to measure
the effective temperature defining flux integral directly provided
that the reddening and angular diameter are known.  We call
this method the flux integration method (FIM).

A straightforward application of the method is to assume a fixed value
of interstellar reddening $E(B-V)$. In this case we can determine
angular diameter and \teff\ by a simple iterative procedure. We
deredden the observed SED using the reddening law of \citet{Gordon03},
a reddening $E(B-V)$, and an initial estimate of \teff, for instance
the value obtained from the TiO fit. We then compare the observed flux
with the model atmosphere flux at a particular wavelength (see below),
and using the initial \teff\ estimate we then calculate the angular
diameter. With this angular diameter we can then turn the observed
flux at all wavelengths into the stellar surface flux, calculate the
flux integral, and hence a new effective temperature. The process is
then iterated until the input and output temperatures converge, which
usually occurs within four to five iterations.

We choose to normalise the model and observed fluxes using the
wavelength range 2.17-2.20\um, for the following reasons. Firstly,
reddening at these wavelengths is minimal. Secondly, at these
wavelengths we are close to the Rayleigh-Jeans domain of the
SED. Here, one can analytically show that the process {\it must}
converge as long as the temperature dependence of the model atmosphere
flux at the wavelength $\lambda_{0}$ of the normalization
$F_{\lambda_{0}}^{model} \propto T_{\rm eff}^{x}$ goes with an
exponent x smaller than four. Finally, the fact that the molecular
line opacity at this wavelength range is weak reduces the uncertainty
of the model atmosphere fluxes, and of the height in the atmosphere at
which the molecules form \citep{Perrin04}. Moreover, at these
wavelengths the line and continuum form at roughly the same depth in
the atmosphere as that which defines the photospheric radius in the
MARCS model atmospheres.

The advantage of this method is that is only very weakly model
dependent through the iterative estimate of the angular diameter at
the wavelength $\lambda_{0}$ and that it uses directly the definition
of $T_{\rm eff}$. Finally, we note that this method is not new. A
similar approach has been applied for hot stars \citep[see for
  instance ][]{R-L82} and also for cool stars \citep{B-S77}.



of 


\begin{table}
  \caption{Results from the TiO and SED methods of estimating \Teff.}
  \setlength{\extrarowheight}{6pt}
  \begin{tabular}{ccccc}
\hline \hline
Star  & $T_{\rm eff}$ (TiO) & $A_{V}$ (TiO) & $T_{\rm eff}$ (SED) & $A_{V}$ 
(SED) \\ 
  & (K) &   & (K) &  \\ 
\hline
SMC 011709 & 3740$^{+140}_{-60}$ & 0.16$^{+0.23}_{-0.16}$ & 4140$^{+50}
_{-50}$ & 0.62$^{+0.10}_{-0.10}$ \\ 
SMC 013740 & 3800$^{+260}_{-80}$ & 0.39$^{+0.38}_{-0.16}$ & 3920$^{+50}
_{-50}$ & 0.52$^{+0.10}_{-0.21}$ \\ 
SMC 020133 & 3640$^{+50}_{-60}$ & 0.47$^{+0.15}_{-0.16}$ & 4000$^{+50}
_{-50}$ & 0.93$^{+0.10}_{-0.20}$ \\ 
SMC 021362 & 3720$^{+80}_{-60}$ & 0.31$^{+0.16}_{-0.15}$ & 3900$^{+50}
_{-50}$ & 0.31$^{+0.10}_{-0.10}$ \\ 
SMC 030616 & 3660$^{+50}_{-60}$ & 0.23$^{+0.08}_{-0.23}$ & 4100$^{+50}
_{-60}$ & 0.62$^{+0.10}_{-0.10}$ \\ 
SMC 034158 & 3760$^{+120}_{-80}$ & 0.31$^{+0.23}_{-0.15}$ & 4100$^{+50}
_{-50}$ & 0.41$^{+0.11}_{-0.20}$ \\ 
SMC 035445 & 3860$^{+520}_{-100}$ & 0.16$^{+0.61}_{-0.16}$ & 4080$^{+50}
_{-50}$ & 0.41$^{+0.11}_{-0.10}$ \\ 
SMC 049478 & 3580$^{+60}_{-50}$ & 0.16$^{+0.23}_{-0.16}$ & 4120$^{+60}
_{-50}$ & 0.72$^{+0.11}_{-0.10}$ \\ 
SMC 050840 & 3640$^{+50}_{-60}$ & 0.31$^{+0.16}_{-0.15}$ & 3940$^{+50}
_{-60}$ & 0.52$^{+0.10}_{-0.21}$ \\ 
SMC 057386 & 3920$^{+460}_{-100}$ & 0.08$^{+0.62}_{-0.08}$ & 4040$^{+50}
_{-50}$ & 0.21$^{+0.10}_{-0.21}$ \\ 
LMC 064048 & 3520$^{+50}_{-60}$ & 0.70$^{+0.23}_{-0.23}$ & 3860$^{+60}
_{-60}$ & 0.62$^{+0.21}_{-0.20}$ \\ 
LMC 067982 & 3580$^{+60}_{-50}$ & 0.47$^{+0.23}_{-0.16}$ & 4180$^{+50}
_{-80}$ & 1.24$^{+0.10}_{-0.21}$ \\ 
LMC 116895 & 3560$^{+50}_{-50}$ & 0.39$^{+0.15}_{-0.23}$ & 4180$^{+60}
_{-100}$ & 0.93$^{+0.10}_{-0.31}$ \\ 
LMC 131735 & 3960$^{+320}_{-120}$ & 0.16$^{+0.46}_{-0.16}$ & 4360$^{+50}
_{-60}$ & 0.62$^{+0.10}_{-0.10}$ \\ 
LMC 136042 & 3460$^{+50}_{-50}$ & 0.23$^{+0.31}_{-0.15}$ & 4200$^{+100}
_{-50}$ & 1.65$^{+0.21}_{-0.10}$ \\ 
LMC 137818 & 3480$^{+60}_{-50}$ & 1.08$^{+0.24}_{-0.23}$ & 4020$^{+60}
_{-50}$ & 0.83$^{+0.10}_{-0.21}$ \\ 
LMC 142202 & 3580$^{+60}_{-50}$ & 0.23$^{+0.31}_{-0.07}$ & 4200$^{+50}
_{-100}$ & 1.24$^{+0.10}_{-0.21}$ \\ 
LMC 143877 & 3660$^{+60}_{-50}$ & 0.23$^{+0.16}_{-0.15}$ & 4360$^{+50}
_{-60}$ & 0.93$^{+0.10}_{-0.10}$ \\ 
LMC 158317 & 3640$^{+60}_{-60}$ & 0.70$^{+0.15}_{-0.23}$ & 4160$^{+60}
_{-80}$ & 1.24$^{+0.10}_{-0.21}$ \\ 
\hline
  \end{tabular}
  \label{tab:teff}
\end{table}

\section{Results} \label{sec:results}

\subsection{TiO temperatures versus SED temperatures}
In \fig{fig:example} we show examples of the fits from the TiO and SED
techniques to two of the stars in the sample. Typically, the TiO fit
is excellent in the region of the TiO bands (left panels). The fitted
temperatures and extinctions are in most cases consistent to within
$\pm$100K or $\pm$0.1mag with those obtained by \citet{Levesque06},
who used similar models with `by-eye' fitting methods. Our errors are
asymmetric due to the non-linear relation between model temperature
and TiO strength, while for higher TiO temperatures the errors tend 
to be larger, due to the diagnostic TiO bands becoming weaker. The SMC 
stars have less TiO absorption for a given temperature due to the reduced 
metal abundance. This effect explains the average spectral type of RSGs 
shifting to earlier subtypes in lower metallicity environments 
\citep{Elias85}.

Longward of $\sim$800nm, however, the fit is poor (right
panels). Specifically, the model with the \Teff\ of the TiO fit
overpredicts the flux at wavelengths greater than $\sim$900nm, as well
as a rise in flux in the $H$-band at $\sim$1.6\um, which we call the
`{\it H-hump}'. It is caused by a minimum in the \Hminus\ opacity, but
is suppressed at high metallicities by molecular absorption, mostly
CO. We do not see the H-hump feature in our XSHOOTER data. This
discrepancy is not an artifact of the fitted extinction: the grey
lines in the right panels of \fig{fig:example} show models with the
temperatures from the TiO fits but which have had the extinction tuned
to give the best fit to the whole SED. Clearly, the fit to the NIR
spectrum is still poor, due to the excess of flux around the H-hump. 

In the case of the SED fits, models with higher values of \Teff\ are
required to fit the NIR continuum, in which the contrast of the H-hump
is reduced. This then in turn requires higher values of \AV. While
the fits to the {\it IJHK} region are good, the {\it BVR}
region is also well matched in most cases, apart from the spectral
range where the TiO features dominate ($\sim$600-800nm). This is
despite the fact that no part of the spectra blueward of 900nm was
used to tune the SED fits. Only in one object is there a substantial
discrepancy between the SED fit and the blue flux (LMC~137818).


\begin{figure}
  \centering
  \includegraphics[bb=20 10 546 516,width=8.5cm]{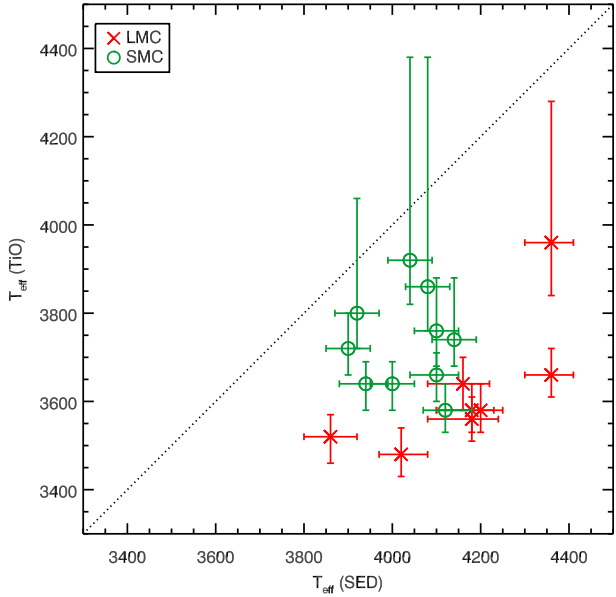}
  \caption{Comparison of the temperatures derived from the TiO band
    fits and from the SED fits. }
 \label{fig:comp}
\end{figure}

\begin{figure}
  \centering
  \includegraphics[bb=0 0 566 425,width=8.5cm]{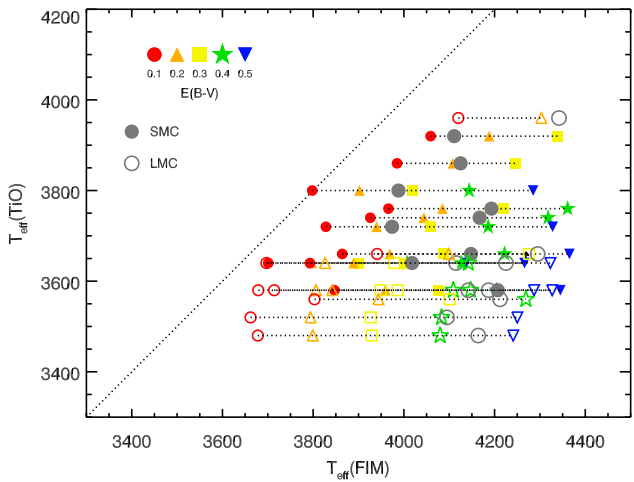}
  \caption{Comparison of the TiO temperatures and those from the flux
    integration method (FIM). The different coloured symbols represent
  the FIM temperature when different values of reddening were
  used. The grey points show the FIM fit with the lowest $\chi^2$
  value (see text for details).}
  \label{fig:FIM}
\end{figure}

\begin{figure*}
  \centering
  \includegraphics[bb=20 20 700 510,width=17cm]{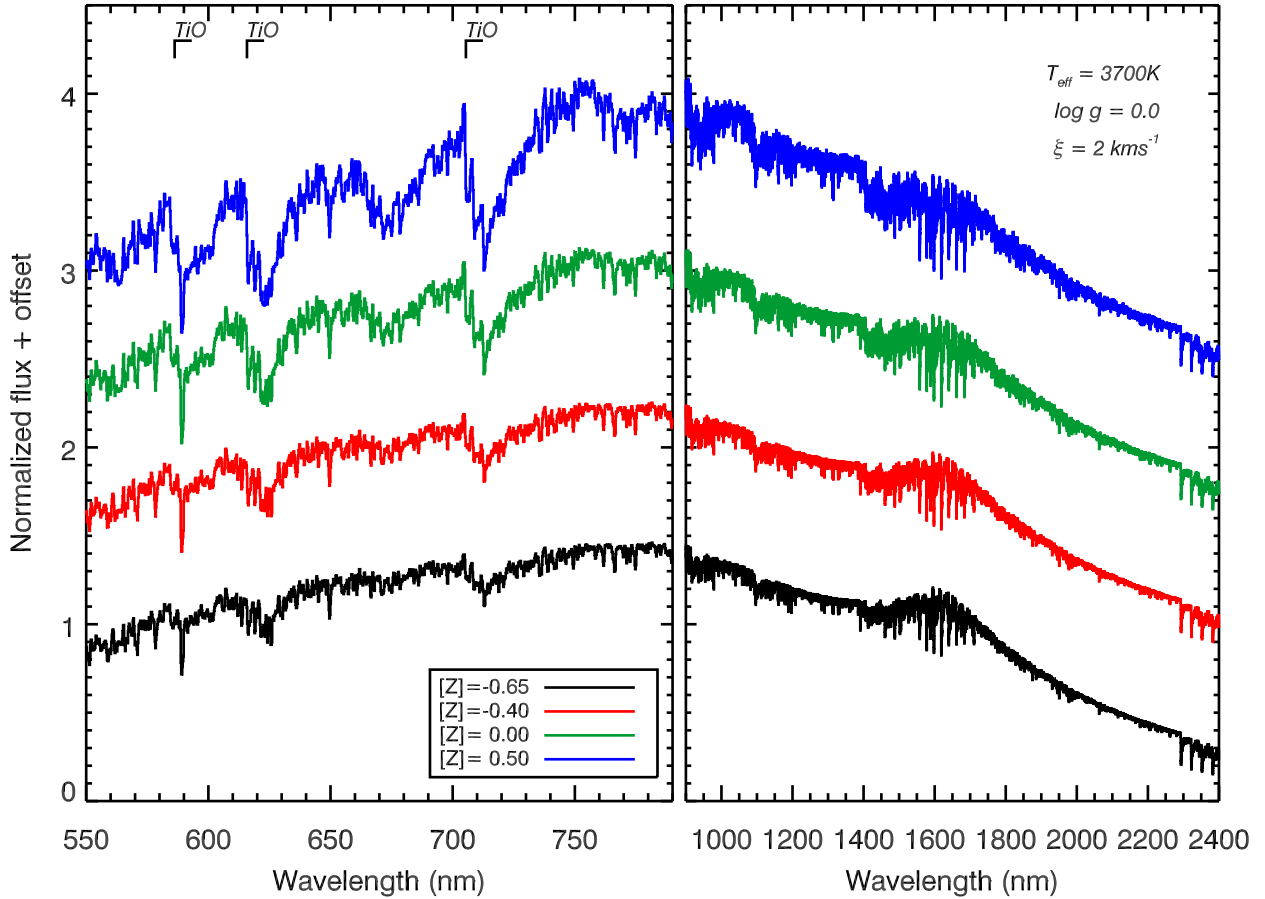}
  \caption{Comparison of the MARCS synthetic spectra in the regions of
    the TiO bands (left) and the H-hump around
    1.5-1.6\um\ (right). All models have \Teff=3700K, \logg=0.0, and
    $\xi$=2\kms. The metallicities of each model are displayed in
    the figure legend.}
  \label{fig:logz_comp}
\end{figure*}

To check whether the H-hump issue is an artifact of poor flux
calibration, we compared the 2MASS photometry of each object with our
synthetic photometry computed by convolving the XSHOOTER spectra with
the 2MASS filter profiles. No object has any systematic offset for the
$H$-band flux compared with $J$ and $K_{S}$, and the two sets of
photometry agree to within the errors. Some objects do have 3$\sigma$
differences in all NIR bands, though these differences are not
correlated with any stellar property. We attribute these differences
to photometric variablity of some of the stars in our sample.

The differences in the \Teff\ values from the two fitting methods are
illustrated in \fig{fig:comp}. A systematic difference in temperatures
is clearly seen. In the case of the TiO fits, a systematic offset is
seen between the average temperatures of the stars in the two
galaxies. For the SED fits, the average \Teff\ values are roughly the
same for both galaxies within the errors, 4170$\pm$170\,K and
4030$\pm$90\,K for the LMC and SMC respectively. The offset of the LMC
and SMC stars in \fig{fig:comp} is due entirely to the metallicity
dependence of the TiO absorption strengths. 

We note that the differences in the \Teff\ and \AV\ values from the
two fits cannot be reconciled by changing the model metallicities,
gravities or microturbulence parameters within sensible boundaries. It
is possible to obtain similar results for the two analysis methods
only if very high metallicities are used (i.e. [Z]$>$0.0 for both LMC
and SMC), but there is little motivation for such high abundances in
the Magellanic Clouds \citep[e.g.][]{Trundle07}. This will be
discussed further in Sect. \ref{sec:reconcile}. 

As well as the SED \Teff\ values being warmer than the TiO fits, the
\AV\ values are also higher (see Table \ref{tab:teff}). It is
reasonable to ask whether these higher extinctions are believable,
since unrealistically high values of \AV\ would put the SED
\Teff\ values in doubt. In fact, these values of \AV\ are consistent
with that of neighbouring stars, plus an extra few tenths of a
magnitude for circumstellar material, consistent observations of
Galactic objects (see Sect.\ \ref{sec:which} for further discussion).



\subsection{Results from the flux integration method}
In this method, the extinction \AV\ is unconstrained. Applying a
reasonable minimum average reddening value towards the Magellanic
Clouds of E(B-V) = 0.1 mag \citep{Zaritsky02,Zaritsky04} leads to a
first result which is displayed as the red circles in
\fig{fig:FIM}. It appears that the temperatures obtained with the
FIM-method \textit{for this minimal value of reddening} are on average 200K
higher than the TiO values. Note that we use the reddening laws
determined by \citet{Gordon03}, which relate extinction $A_{V}$ and
reddening through $A_{V} = R_{V}E(B-V)$ with $R_{V}$ equal to 2.7 and
3.4 in the LMC and SMC, respectively.

Of course, as already indicated by the results obtained with the
SED-fitting method it is very likely the reddening and extinction
towards the sites of the RSGs are much higher than the minimal
reddening. To account for possible reddening values different from
0.1mag we assume a range of reddening values between \ebmv=0.0mag and
0.5mag and determine \Teff\ in the same way as above at each reddening
(orange, yellow, green and blue symbols in \fig{fig:FIM}). In
addition, we compare the dereddened observed fluxes with the model
fluxes at three wavelength bands which are free of molecular
absorption (1.605-1.702\um, 1.212-1.278\um, and 1.000-1.080\um) and
calculate $\chi^{2}$ values for each \ebmv. The minimum of $\chi^{2}$
is then used to determine the best-fitting \ebmv\ and \Teff\ (grey
points in \fig{fig:FIM}). We note that this extended method aiming at
a simultaneous determination of \Teff\ and \ebmv\ also depends on the
use of model atmospheres, and is very similar to the SED method. As
such, these best-fits cannot be considered to be entirely independent
of the results from the SED fits. 

What these results {\it do} show is that, even for minimal reddening,
the FIM temperatures are always higher than the TiO temperatures by
100-200K. Increasing the reddening only serves to increase the
discrepancy further. We again state that deviations from the 
standard interstellar extinction law cannot explain this discrepancy, since 
this can only account for temperature differences of $\pm$50K even for the most extreme values of $R_V$.  
This served to cast further doubt on the
reliability of the TiO temperatures.



\begin{figure}
  \centering
  \includegraphics[bb=0 0 556 500,width=8.5cm]{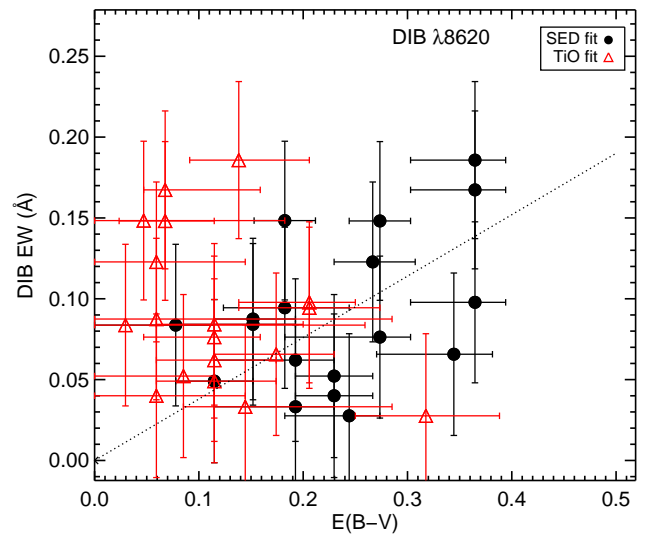}
  \caption{The amount of reddening deduced from each of the TiO band
    and SED fits, versus the equivalent width of the DIB at
    8620\AA. The dotted line shows the expected relation in the
    Galaxy, from \citet{Munari08}}
  \label{fig:DIB}
\end{figure}

\section{Discussion}  \label{sec:disc}

\subsection{Which temperature scale is correct?} \label{sec:which}
Since the different fitting methods give two different temperatures
for the stars observed, it is natural to ask which, if any, is
correct. One major reason to be suspicious of the TiO temperatures is
that they consistently overpredict the flux at wavelengths $>$900nm,
dramatically so in some cases (e.g.\ LMC~064648, LMC~137818, see
\fig{fig:example}). Though the SED fits overpredict the {\it optical}
flux, this may be plausibly explained by TiO absorption in the MARCS
models (see later). Conversely, there is no obvious physical mechanism
that would cause us to greatly overestimate the NIR flux whilst
correctly matching that in the optical. One cannot reduce the
extinction to rebalance the optical and NIR fluxes as the \AV\ values
from the TiO fits are already very low. We note that this discrepancy
was already flagged in L06, who found that the $V-K$ colours of RSGs
did not match the predictions of MARCS models with the same \teff.

One other strong piece of evidence that the TiO temperatures are too
low comes from the FIM results. This method, which uses the observed
flux integrated over all wavelengths, is insensitive to model
dependent features such as the predicted strengths of molecular
absorption. Instead, it assumes only that the models correctly predict
the ratios of the total flux to that in a specified wavelength
interval, which we chose to be a small section of the $K$-band. These
results showed that even for very low values of reddening the FIM
temperatures were systematically higher than the TiO temperatures.

Another line of investigation is to examine the \AV\ implied by each
fitting method. Here, we consider only the TiO and SED results, since
the FIM results are not entirely independent of the latter. For a
given fit, to some extent there will always be a degree of degeneracy
between \Teff\ and \AV, since the flux peak can be shifted for example
to the blue by either increasing \Teff\ or decreasing \AV \footnote{This 
applies mainly to the SED method, though it is also relevant to the TiO 
method for earlier spectral types, i.e. when the TiO bands are weak.}. As the 
two methods (TiO and SED) place emphasis on different spectral properties,
each gives a different \AV. Therefore, by finding some other
diagnostic of \AV, we may assess which method is providing the most
reliable results.

An independent measurement of extinction may be obtained from the
strengths of the diffuse interstellar bands (DIBs). Many DIBs are
known to strongly correlate with reddening
\citep[e.g.][]{Cox06,Cox07}; however the majority of these features
are located in the $V$ and $R$ bands, meaning that in RSGs they are
blended with many intrinsic features. One DIB located in a relatively
clean spectral region is that at 8620\AA, identified in the spectra of
stars in the RAdial Velocity Experiment (RAVE) sample
\citep{Munari08}. In \fig{fig:DIB} we plot the strength of this line
from each star in our sample against the reddening $E(B-V)$ obtained
from each of the fitting methods. We find that the reddenings from 
the TiO fits appear to be poorly correlated with the DIB strengths. The 
formal Pearson coefficient is small and negative (-0.35), implying that if 
anything the data are \textit{anticorrelated}. There is a greater degree of 
correlation however between the DIB strengths and the reddenings from the 
SED fits. This time the Pearson coefficient is positive, though still quite 
low (+0.44). A linear fit to the data using the GSFC IDL routine 
\textsc{fitexy}, which considers errors in both variables\footnote{The 
routine assumes the errors are symmetric, whereas the errors in our E(B-V) 
values tend to be asymmetric. However, since the level of asymmetry is 
small, and as we are not interested in a precise formal fit to the data, we 
used the larger of the two errors in this analysis.}, yields a slope and 
offset consistent with the fit to the RAVE stars within the errors (the 
latter shown as a dashed line in \fig{fig:DIB}). While this evidence is far 
from conclusive, it does provide circumstantial evidence in favour of the 
SED \Teff\ values over those from the TiO fits.

\begin{figure}
  \centering
  \includegraphics[width=8.5cm]{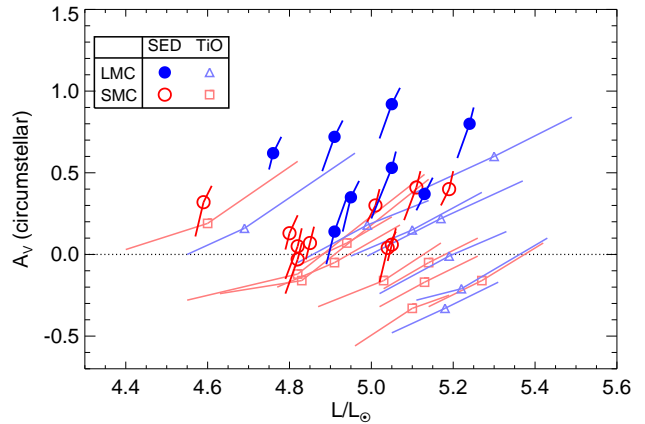}
  \caption{The circumstellar extinction around each star, found by
    subtracting the local from the total extinction, as a function of
    luminosity. The results of both the SED and TiO fits are
    shown. Note the asymmetric and correlated errors on each datapoint.}
  \label{fig:lum-av}
\end{figure}

We can also estimate the extinction towards each star from that of the
surrounding stars. Though RSGs may have some extra circumstellar
extinction caused by their mass-loss history, the surrounding stars
should give us an idea as to the foreground extinction, both Galactic
and internal to the star's host galaxy, and so give an indication as
the the lower limit of the \AV\ towards each object. 

The foreground extinction was measured by looking at that towards
stars within 1\arcmin\ of our targets from
\citet{Zaritsky02,Zaritsky04}. We experimented with several methods of
interpolating the extinction at the location of our targets, such as
binning in right-ascension and declination or in radius from the
target star. In the end we found that simply median averaging the
stars within 0.5\arcmin\ gave results similar to any other method
tried, and was stable to within $\pm$0.2mags in \AV.

In \fig{fig:lum-av} we plot the circumstellar extinction (the total
extinction from the spectral fits minus the foreground extinction from
the neighbouring stars) as a function of stellar luminosity $L$ (see
Sect.\ \ref{sec:stellarevol} for details on how we measure $L$). The
first thing to note is that the circumstellar extinction implied by
the TiO fits is often negative -- that is, the localized extinction is
{\it greater} than the total extinction measured from the TiO
region. In contrast, the total extinction as measured from the SED
fits is always at least equal to that of the local extinction. Some
objects, especially the more luminous ones, show evidence of
substantial circumstellar extinction of up to \AV=1\,mag, consistent
with mid-IR observations of Galactic RSGs \citep[e.g.][]{deWit08},
and expectations calculated from simplified mass-loss histories
\citep[][]{W-E12}. In addition, there is a separation between the LMC
and SMC stars, suggesting a possible metallicity dependence on RSG
wind density.

To summarise, we conclude that the \teff\ measurements from the
strengths of the TiO lines are unreliable, since model fits to these
lines overpredict the NIR flux, and extinctions implied are
anomalously low. The SED temperatures on the other hand provide good
fits to all parts of the spectrum apart from the TiO bands. They also
imply extinctions that are consistent with their neighbouring stars
with a contribution from a circumstellar component. In the following
section we discuss how the mismatch between the SED and TiO results
may arise, and discuss possible solutions.

\subsection{Reconcilling the two
  \Teff\ measurements} \label{sec:reconcile} 
Given this clear discrepancy between the different independent
temperature measurements, we now explore various explanations for this
result. We first note that the differences in the \Teff\ and \AV\ values
from the two fits cannot be reconciled by changing the model gravities
or microturbulence parameters within sensible boundaries.

The discrepancy between the two methods may disappear at higher
metallicities. In \fig{fig:logz_comp} we show the MARCS model spectra
at fixed \Teff, \logg\ and $\xi$, for SMC-like, LMC-like, Solar, and
Solar+0.5dex metallicities. The strengths of the TiO bands are
strongly correlated with [Z], as one would expect. In addition, the
strength of the H-hump is inversely correlated with [Z], as atomic and
molecular absorption eat away at the sub-peak caused by a minimum in
the \hminus\ opacity. These two effects may therefore act to bring the
two \Teff\ measurements together: increased TiO absorption would
require models with higher \Teff\ values to provide a match to TiO
bands of a given strength; while a reduced H-hump would mean that good
SED fits would be obtained at lower \Teff\ values. Therefore,
potentially one could obtain similar results for the two analysis
methods if very high metallicities are used (i.e. [Z]$>$0.0 for both
LMC and SMC). However there is little motivation for such high
abundances in the Magellanic Clouds. Young-age metallicity tracers
such as cepheids, F~supergiants and B~stars consistently show average
abundances of \hbox{[Z]$_{\rm LMC}$ = -0.4} and \hbox{[Z]$_{\rm LMC}$
  = -0.6}, with errors on the means of $\pm 0.05$dex, and typical
intrasample dispersions of less than $\pm 0.02$dex
\citep[e.g.][]{Andrievsky01,Romaniello05,Keller-Wood06,Trundle07}.

\begin{figure*}
  \centering
  \includegraphics[bb=-10 0 860 340,width=17cm]
{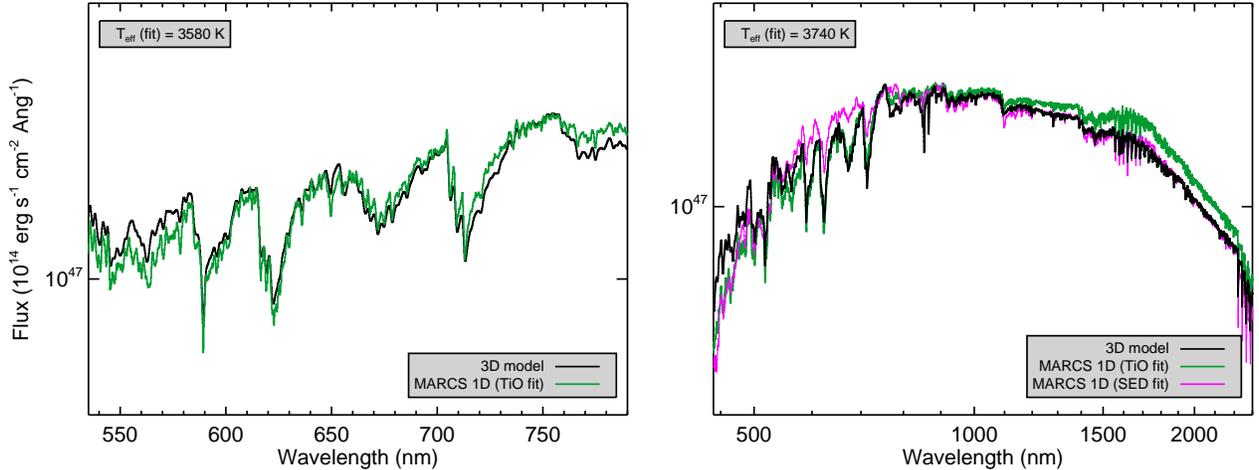}
  \caption{Analysis of the 3D {\sc co$^{5}$bold} RSG spectrum of
    \citet{Chiavassa11} with the 1D MARCS models. The left panel shows
  the 3D spectrum (black) along with the fit to the TiO bands
  (green), yielding a best-fit \Teff=3600K. The right panel shows that
  this fit overpredicts the flux at the $H$-hump, and in the
  $K$-band. Meanwhile, a fit to the NIR continuum regions (magenta)
  gives a higher temperature of 3800K, while underpredicting the
  strengths of the TiO bands. This mimics the behaviour seen in our
  spectra of RSGs in the Magellanic Clouds (see \fig{fig:example}).}
  \label{fig:3D}
\end{figure*}

\subsubsection{Adjusting CNO abundances}
Rather than changing the adundances of all metals in a uniform way,one
might suggest that adjusting only the CNO mixture to reflect the
products of nuclear burning would alter the appearance of the
spectrum. The relative abundances of TiO, CN, and CO should depend on
the CNO mix, and all are major sources of opacity. Altering the CNO
mixture may therefore alter the temperature structure of the star,
changing the strengths of the diagnostic absorption features. 

To investigate this, we computed a model with \Teff=3700K, [Z]=-0.5,
\logg=0.0, and $\xi$=2\kms, with a CNO mixture altered to reflect
stellar evolution. Specifically, we reduced C by 0.4dex, and
correspondingly increased N by +0.54dex. This is similar to the
surface abundances of Galactic RSGs \citep{RSGCabund}. We then
compared this to a model with the same parameters other than the CNO
mix set to Solar. The evolved CNO model had slightly weaker CO and CN
features, due to the reduced C abundance; while TiO absorption was
slightly increased due to less O being tied up in CO. However, the
differences between the two were minimal. Analysing the evolved CNO
model in the same way as the XSHOOTER spectra in this paper, we found
that this led to \Teff\ discrepancies of less than 20K. We therefore
conclude that deviations from a Solar CNO mixture cannot explain the
results presented here.


%

\begin{figure}
  \centering
  \includegraphics[bb=0 0 623 566,width=8.5cm]{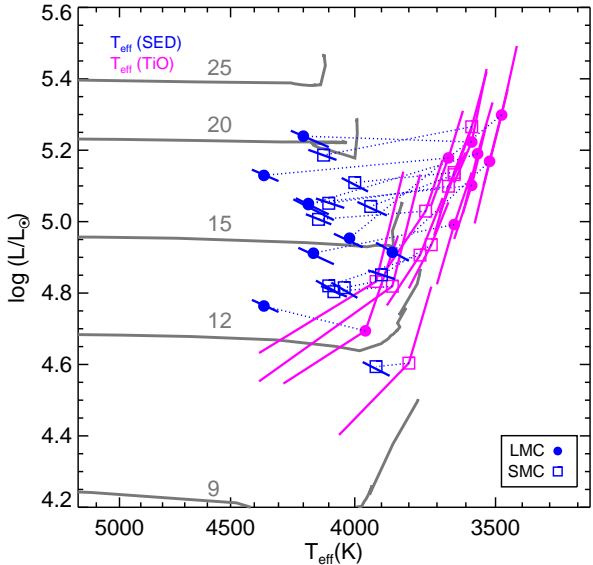}
  \caption{HR-diagram of the stars in our sample. The magenta points
    show the locations of the stars when the TiO temperatures and
    extinctions are used, while the blue points use the values derived
    from the SED fits. Note that the errors on each quantity are
    correlated and asymmetric. The grey lines show the Geneva mass
    tracks for stars with initial rotation 400\kms\ and LMC-like
    metallicity.}
  \label{fig:HR}
\end{figure}

\subsubsection{Convection, granulation, and temperature structure}


As the strengths of the TiO bands are very temperature sensitive,
inadequacies in the temperature structure of the models may lead to a
discrepancy in the absorption strength as a function of \teff, giving
rise to the effects described in this paper. Indeed, this was
suggested in L06 as an explanation for the $V-K$ discrepancy. Here we
discuss possible causes of such an effect.

Convection is known to occur in the envelopes of
RSGs, and is the physical justification for the inclusion of both
micro- and macro-turbulence in spectral synthesis models. 
In order to self-consistently model these effects rather than
including them in an ad-hoc fashion, one needs to switch from 1-D
hydrostatic to 3-D hydrodynamic stellar models. Such work is still in
its early stages, but \citet{Chiavassa11} present a 3-D non-gray model
of a RSG, along with a synthesized spectrum, and comparisons to 1-D
models. These authors show that the temperature structure for the 3-D
models is subtley different from that of the 1-D model, resulting in
the formation zone of the TiO lines shifting outwards to larger radii
and lower temperatures. In terms of the time-averaged SED of the 3-D model, they
found that their non-grey model had similar TiO band strength to the
1-D model with a low \Teff=3400K, but that the SED $>$1\um\ looked
more like the 1-D model with \Teff=3700K \footnote{
  \citet{Chiavassa11} quote 3430K as the temperature of their
  model. However, this is the result of a complex non-standard
  averaging procedure. The integrated flux of their model corresponds
  to \Teff=3700K, which is also that model's temperature at $\tau=1$.}.

In \fig{fig:3D} we show this quantitatively, by fitting the Chiavassa
et al.\ $<$3-D$>$ spectrum with 1D MARCS models just as in our analysis of
the XSHOOTER spectra of the LMC/SMC stars. We used the same gravity
and metallicity as were computed for the 3-D model (\logg=-0.34,
[Z]=0.0)\footnote{In their paper, \citet{Chiavassa11} state that due
  to the effect of convective pressure the {\it effective} gravity is
  \logg=-0.65. However, we found that adjusting the gravity had little
  effect on our fits, other than a change in the opacity below
  500nm.}. As with our observations, we find that the TiO fit has too
much flux in the NIR, while the fitted temperature is lower than the
input value ($T_{\rm TiO} = 3580 \pm 50 K$). The SED fit again
provides an excellent match to all regions 500-2500nm, other than in
the regions of the TiO bands, and has a temperature closer to the
integrated flux of the $<$3-D$>$ model ($T_{\rm SED} = 3740 \pm 50 K$). Both
fitted models are discrepant at short wavelengths ($<$500nm).

Though a detailed analysis of 3-D model spectra is beyond the scope of
this current work, it seems as though 3-D effects may serve to
reconcile the discrepancy in the \Teff\ as measured from different
regions of the spectrum we have highlighted here. In particular, the
temperature in the outer layers of the 3-D model where the TiO lines
form is somewhat lower, leading to increased TiO absorption. However,
the temperature at optical depth $\tau \approx 1$, where the continuum
forms, is roughly the same as in the 1-D model (see their Fig.\ 5,
top-right panel). For this reason we conclude that the temperatures
measured from the continuum regions are more reliable than those
measured from the TiO bands.

Finally, we suggest that another potential complicating factor could
be the difficulty in delineating the stellar photosphere, the
molecular formation zone, and the stellar wind. If molecules form in
the wind, or close to it, then their strengths may depend on wind
density. If the mass-loss rates of RSGs are a function of their
luminosity \citep{Bonanos10}, then the strengths of the molecular
absorption bands would be highly sensitive to luminosity as well as
temperature. If mass-loss rates are also [Z] dependent \citep[][ see
  also \fig{fig:lum-av} in this paper]{M-J11}, then this may serve to
increase this sensitivity of the band strengths to luminosity.


\subsection{Comparison with Galactic RSGs}

Since there is a known dependence of the TiO band strengths on
metallicity, it is interesting to ask how a sample of Galactic RSGs
would compare to the LMC and SMC stars studied here. However, obtaining data 
comparable to that presented here is problematic. The most well-known 
Galactic RSGs, such as $\mu$~Cep and $\alpha$~Ori, are are too bright to 
observe with VLT/XSHOOTER, which currently is the only instrument capable of simultaneous optical/near-IR spectrophotometry. Meanwhile, more distant RSGs such as those in the Scutum RSG clusters at 6.6kpc \citep[e.g.][]{Figer06,RSGC2paper} have large foreground extinction of \AV$\ga$15\,mag, making optical spectrophotometry of the TiO bands extremely challenging.

We have obtained NIR spectroscopy of stars in Per~OB1 with IRFT+SpeX (Gazak 
et al., in prep), and combined these data with the optical spectroscopy of
L05. We then analysed these stars in the same way as our XSHOOTER
data. We found very similar results, specifically that the SED
temperatures were consistently 4100$\pm$150\,K, whereas the TiO
temperatures were much cooler and varied with spectral type (agreeing
well with the results of L05). 

However, there are potential causes for concern with these
data. Firstly they are non-contemporaneous, the optical and NIR
spectroscopy taken several years apart. Secondly, the SpeX slit was
$<$1\arcsec\ wide, leading to inevitable slit losses and compromising
the accuracy of the spectrophotometry. Finally, the absolute
photometric calibration was done with 2MASS, in which the stars are
saturated and have photometric errors of $\pm$0.2mags. Since the data
quality does not match that of our excellent XSHOOTER data, we have
chosen not to present the results of that analysis here.

\subsection{Implications for stellar evolution} \label{sec:stellarevol}
To see how the difference in \Teff\ affects the stars' positions in
the H-R diagram, we determine the luminosities of the stars for the
TiO and SED \Teff\ estimates. Here, we use distance moduli of 18.47
and 18.95 for the LMC and SMC respectively (averages of a number of
measurements, taken from the NASA Extragalactic Database).

For the TiO temperatures we first take the $V$-band magnitude of each
star from our synthetic photometry. We then correct for the
\AV\ determined from the TiO fits, and apply the bolometric
corrections from L06.

For the SED fits, the luminosity was obtained by integrating under the
SED. The flux in the regions of high telluric extinction was
interpolated using the best-fitting MARCS model for a given star. The
flux longward of 2.5\um\ was estimated by logarithmic interpolation
between a star's flux at 2.3\um\ (i.e. before the CO bandhead
absorption kicks in) and the four {\it Spitzer}/IRAC fluxes at 3.6\um,
4.5\um, 5.8\um\ and 8.0\um. The total flux outside our observed region
of 400nm-8.0\um\ was considered negligible. The flux was dereddened
using the \citet{Gordon03} extinction law, and assuming that the
correction for wavelengths greater than 2.5\um\ was effectively
zero. The bolometric flux was then calculated by integrating under the
dereddened SED.

The H-R diagram for the stars studied here is shown in
\fig{fig:HR}. When the temperatures and luminosities are updated to
reflect the results of the SED analyses, we find that the stars move
to warmer temperatures at roughly constant luminosity. The reason that
the luminosities are largly unchanged is that, when analysing the
SEDs, we find that both \Teff\ and \AV\ are higher than the results of
the TiO fits. This means that, in the TiO fits, the smaller extinction
correction is cancelled out by a larger bolometric correction.
 
In terms of the location of the stars in the H-R diagram in relation
to the predictions of stellar evolutionary theory, the SED
temperatures bring the RSGs into much better agreement with the Geneva
rotating model calculations. It was pointed out by L06 that the TiO
temperatures were slightly cooler than the coolest point (i.e.\ the
Hayashi limit) of the Geneva stellar tracks. We note however that the
other contemporary stellar evolution codes, such as the Bonn or
Cambridge codes \citep{Eldridge08,Brott11} predict that the Hayashi
limit at LMC metallicity should be cooler than that of the Geneva
models, at around 3700K. It is therefore worth discussing the nature
of the Hayashi limit of RSGs a little further.

The Hayashi limit is the maximum size to which a star can grow whilst
remaining in hydrostatic equilibrium, and depends on a number of
factors, but is mainly governed by the opacity of the envelope
(i.e. metallicity) and the convective mixing length parameter $\alpha$
\citep[e.g.][]{Lucy86}. In most stellar codes, $\alpha$ is tuned to
reproduce the observed properties of the Sun, and is assumed to be the
same for stars of all masses and metallicities. Some authors have
argued that a single value of $\alpha$ cannot reproduce the
colour-magnitude diagrams of globular clusters \citep{Ferraro06},
while there is some debate about whether $\alpha$ may be dependent on
metallicity \citep{Salaris-Cassisi96,Keller99,Palmieri02}.

Given this discussion, it is perhaps surprising that though the Bonn,
Cambridge and Geneva codes all use the same mixing length, they
predict different temperatures for the Hayashi limit of RSGs. The
reasons for this are unclear. The solution to this problem will
ultimately come from reliable temperature measurements of RSGs. The
results presented here are therefore a significant step towards
understanding the nature of convection in cool stars. 


\begin{figure}
  \centering
  \includegraphics[bb=0 0 576 420,width=8.5cm]{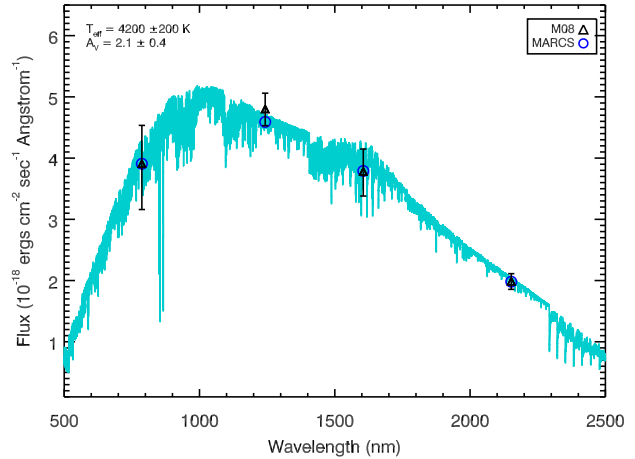}
  \caption{The pre-explosion photometry of SN2008bk's progenitor, and
    the best-fit model with \Teff=4200K. The photometry from
    \citet{Mattila08} is represented by the black triangles, while the
    synthetic photometry from the best-fitting MARCS model is shown by
    the blue circles. The full SED of that model is shown by the cyan
    line. }
  \label{fig:SN2008bk}
\end{figure}

\begin{table}
\centering
  \caption{Mean bolometric corrections and standard deviations for
    RSGs in each of the two Magellanic Clouds. }
  \label{tab:BC}
  \begin{tabular}{lcc}
    \hline \hline
    Galaxy & LMC & SMC \\
    \hline
    {\it BC$_{K}$} &  2.69 $\pm$ 0.12 &  2.69 $\pm$ 0.06 \\
    \hline
  \end{tabular}
\end{table}

\begin{table}
  \centering
  \caption{Parameters for use with Eq.\ (\ref{equ:lcalib}) to derive
    bolometric luminosities from single-band photometry of RSGs. }
  \label{tab:lcalib}
  \begin{tabular}{lccc}
\hline \hline
Band & $a$ & $b$ & $\sigma \log(L)$ \\
\hline
V  &  3.12 $\pm$ 0.06  &  -0.29 $\pm$ 0.01  &  0.12 \\
R  &  2.44 $\pm$ 0.07  &  -0.34 $\pm$ 0.01  &  0.09 \\
I  &  1.90 $\pm$ 0.08  &  -0.37 $\pm$ 0.01  &  0.06 \\
J  &  1.30 $\pm$ 0.09  &  -0.39 $\pm$ 0.01  &  0.03 \\
H  &  0.97 $\pm$ 0.10  &  -0.40 $\pm$ 0.01  &  0.04 \\
K  &  0.90 $\pm$ 0.11  &  -0.40 $\pm$ 0.01  &  0.04 \\
\hline
  \end{tabular}
\end{table}

\subsection{Bolometric corrections, luminosity calibrations, and
  application to supernova progenitors} 

\subsubsection{Bolometric corrections}

The uncertainties that we highlight in the RSG temperature scale are
extremely important for the study of SN progenitors. Object
brightnesses and colours in pre-explosion images are used to place the
progenitor on the HR~diagram, and by comparison with stellar models
determine the progenitor mass. Clearly, such work is sensitive to both
the progenitor object's extinction and, through the bolometric
correction, the calibration of colour with effective temperature.

We calculate the BC for each star in our sample from the bolometric
flux, as determined from the SED fits in Sect.\ \ref{sec:stellarevol},
the flux from the synthetic photometry (Sect.\ \ref{sec:synthphot})
once corrected for extinction (from the SED fits), and assuming an
absolute bolometric magnitude for the Sun of 4.77.

We find that, as predicted, the BCs at various bands are correlated
with temperature. However, we have shown here that the temperatures of
RSGs, as determined from the continuum regions of the SED, are uniform
to within $\pm$100K. This means that the amplitude of variations in BC
are small, particularly in the $K$-band where we are close to the flux
peak and where there is little contribution from molecular
absorption. Hence, to a good approximation we can assume that for RSGs
BC$_K$ is constant to within $\sim$0.1\,mag.

In Table \ref{tab:BC} we list the average BCs for the RSGs in both the
LMC and SMC in the $K_{S}$ band. The BCs for both galaxies are
consistent with one another to within the errors. This has the
important implication that we can determine the bolometric apparent
magnitude of a RSG -- and if the distance is known, the luminosity --
from only one photometric data-point.

As is to be expected, the BC$_K$ values presented here are much
changed compared to those quoted in L06. This is caused by the
incorrect assumption that the models which best-fit the TiO bands will
accurately predict the flux at all wavelengths. In fact, the LMC
BC$_K$ scale is closer to that of \citet{Elias85}, which relied on
averaged instrinsic colours for M supergiants and trapezoidal
integration of optical/near-IR broad-band photometry. Our results for
the SMC however are greatly different to those of Elias et al., since
those authors did not have accurate intrinsic colours for SMC RSGs.

\subsubsection{Luminosity calibrations}

To be able to determine the luminosity of the star from a given
photometric point, we must first measure and subtract the
extinction. We separate this extinction into two components,
interstellar (IS) and circumstellar (CS). The IS extinction is
measured from the surrounding stars (Sect.\ \ref{sec:which}),
whereas the total (IS+CS) is measured from the SED fits. Since the CS
component is typically small (\AV$\la$1), in the infrared bands this
correction will be minor. Therefore, if the IS component can be
estimated, one can estimate the luminosity of a RSG from $K$-band
photometry alone. Furthermore, since the CS extinction is somewhat
correlated with luminosity (see Sect.\ \ref{sec:which}), it is
possible to calibrate this effect out and estimate the luminosity from
any pass-band.

To do this, we plotted the magnitudes of each star though a given
pass-band, corrected for foreground extinction, against the
luminosities from the SED fits. We then used the IDL procedure {\sc
  linfit} to determine the coefficients of the linear fits $a$ and $b$
to these relations, which we list in Table \ref{tab:lcalib}. The
bolometric luminosity of a RSG can then be determined from,

\begin{equation}
log(L/L_{\odot}) = a + b (m_{\lambda} - \mu)
\label{equ:lcalib}
\end{equation}

\noindent where $m_{\lambda}$ is the apparent magnitude at a given
pass-band {\it after} correction for interstellar extinction, and
$\mu$ is the distance modulus. In the cases of {\it IJHK}, $b$ is
close to 0.4(=2.5$^{-1}$), which follows simply from the magnitude
definition. At $V$ and $R$ bands, $b$ begins to diverge from this
value, which is due to the flux at these bands being sensitive to CS
extinction and TiO absorption, both of which are luminosity
dependent. The final column of Table \ref{tab:lcalib} shows the
r.m.s. standard deviations of the residuals between the luminosity
from the SED fits and those determined from
Eq.\ (\ref{equ:lcalib}). In practice, the dominant sources of error on
$\log(L/L_{\odot}$ are the uncertainties on $a$ and $b$, which
corresponds to $\sim$0.1dex at $K$ and $\sim$0.20dex at $V$.

We note that we found no appreciable differences in the results when
the sample was divided into stars from each of the LMC and SMC. We
therefore combined all the stars to give greater signal-to-noise. In
future, using larger samples of stars, these relations could be
improved upon by analysing the stars from the two galaxies seperately
to take account of metallicity effects, increasing the accuracy.

\subsubsection{Application to pre-explosion photometry of Type II-P
  supernovae} 

To assess the potential impact on the field of SN progenitors we
re-analyse the pre-explosion photometry of the Type II-P supernova
2008bk using our SED temperatures. This object was chosen as it has
LMC-like metallicity, consistent with the objects studied in this
current work, and pre-explosion photometry in four bands
\citep{Mattila08}, rather than the usual one or two which is the case
for most progenitors \citep{Smartt09}. In their paper, Mattila et
al.\ fit the photometry as an M4 supergiant, \AV=1.0$\pm$0.5, and
using the temperature scale of \citet{Levesque05}, $BC_{K} = 2.9 \pm
0.1$. This led to a bolometric luminosity of $\log (L/L_{\odot}) = 4.6
\pm 0.1$, and by comparison to stellar models, an initial mass for the
progenitor star of $9^{+4}_{-1}$\msun. They note that the SED may also
be fit by a M0 supergiant and a higher extinction of \AV=3.0.

To re-analyse the pre-explosion photometry of SN2008bk, we restrict
the model \Teff\ to between 4130$\pm$150\,K, consistent with our
results for the LMC, and use an LMC-like metallicity of [Z]=-0.4. We
again use \logg=0.0 and $\xi=2.0$\kms, noting that these latter two
parameters make little difference to the results. For each temperature
we determine the \AV\ required to produce the best fit.

We find that the photometry can be fit very well by models in this
temperature range, requiring an extinction of
\AV=2.1$\pm$0.4. Following the analysis procedure of
\citet{Mattila08}, we use the object's $K$-band photometry
($K$=18.34$\pm$0.07), the bolometric correction appropriate for the
LMC ($BC_{K}=2.69\pm0.12$), the extinction at $K$
\citep[=\AV$\times$0.11][]{R-L85}, and a distance modulus of
27.9$\pm$0.2. This gives us a bolometric luminosity for the progenitor
of $\log (L/L_{\odot}) = 4.75 \pm 0.10$, slightly higher than the
\citet{Mattila08} estimate.

We can also estimate \lbol\ using our emprical relations in the
previous section. We use the object's $K$-band photometry, since this
is the least sensitive to extinction. Assuming negligible foreground
extinction, we find $\log (L/L_{\odot}) = 4.72 \pm 0.14$. Mattila et
al.\ suggest that the foreground Galactic extinction to the host
galaxy is low (\AV$\sim$0.06), but that a comparison to a SN with a
similar light-curve suggests that there may be around 0.3mag of
extinction at $V$ internal to that galaxy. This is a negligible amount
of extinction at $K$, and would increase the luminosity of the
progenitor by only $\sim$0.01dex.

To convert this luminosity into an initial stellar mass, we compare to
the stellar evolutionary tracks at LMC metallicity. We use the tracks
of both the Bonn \citep{Brott11} and Geneva groups. Our assumed
temperature for the progenitor star means that its location in the
HR-diagram does not coincide with the terminal point of any of these
tracks, which evolve to temperatures of $\sim$3500K. However, as noted
earlier, this terminal \Teff\ is set by the adopted value for the
mixing length in the stellar structure models, which is poorly
constrained for massive stars. In order to estimate the initial mass
of the progenitor, we follow the methodology of \citet[][ and
  references therein]{Smartt09} and compare the star's luminosity to
the terminal luminosities of the mass tracks, which corresponds to the
luminosity at the end of He burning. In this way, we find a progenitor
mass for SN2008bk of $M = 12^{+2}_{-1}$\msun\ using the Bonn tracks,
and $M = 11\pm1$\msun\ with Geneva tracks. This is $\sim$30\% higher
than the previous estimate of \citet{Mattila08}, though within their
errors.


We note that our extinction is slightly higher than that derived by
\citet{Mattila08}, and that those authors argued against a higher
\AV\ solution due to the lack of signs of circumstellar extinction in
the SN spectrum. However, it is entirely plausible that for a
relatively small amount of circumstellar extinction close to the star,
this material would be destroyed almost instantly by the SN
explosion. Indeed, \citet[][]{Fraser12} make a similar argument for
the progenitor of SN2012aw, for which multi-band pre-explosion
photometry was also available. 

In summary, we find that if we take the temperatures of RSGs as
derived from fits to the near-IR SED, the extinction increases, while
the magnitude of the bolometric correction decreases. Though these two
effects work in opposite directions, in the case of SN2008bk the net
effect is for the progenitor star's inferred bolometric luminosity to
increase by 0.14dex. Therefore, the initial mass of the progenitor is
higher compared to that derived using the
\citet{Levesque05,Levesque06} temperature scale. If the change in
temperature scale affects the properties of progenitors in the same
way, this may in part explain the so-called `Red Supergiant problem'
\citep{Smartt09}, whereby the progenitor masses of Type~II-P SNe tend
to be lower than those expected from initial mass function
arguments. This is similar to the suggestion of \citet{W-E12} that the
`missing' SN progenitors could be RSGs which have sufficient
circumstellar extinction to push them below the detection limits of
the pre-explosion imaging.

\subsection{Concluding remarks on the temperatures of \\ Red Supergiants}
In the course of this work, we have shown that for RSGs the definition
of a characteristic temperature is a non-trivial task. The definition
of effective temperature, $T_{\rm eff} \equiv L/(4 \pi R^2 \sigma)$, whilst
unambiguous for stars such as the Sun with well-defined photospheres,
can be problematic to define for stars with extended
atmospheres. Simply defining the photosphere to be where the optical
depth $\tau = 2/3$ is insufficient, since this value can de reached at
very different radii depending on the observed wavelength. Any
flux-averaged value of $\tau$ will therefore be sensitive radiative
transfer processes in the stellar atmosphere. 

Another problem is finding accurate and robust spectral diagnostics of
\teff. For M supergiants, it has always been assumed that the
strengths of the TiO bands, which define the spectral classification
sequence of these stars, were useful diagnostics of effective
temperature. However, any diagnostic line is sensitive {\it only} to
the local temperature in the layers where the line forms. In the case
of the TiO lines, these form high in the atmosphere, and so are very
sensitive to the radial temperature structure. As we have shown in our
comparison to 3-D hydrodynamic models, two stars with the same
luminosity and radius can have different TiO strengths if the
temperature structure is different. However the continuum, which forms
at much deeper layers, is indistinguishable. Though there is
potentially a degeneracy with reddening, this degeneracy can be broken
by considering the magnitude of the \Hminus\ opacity minimum. We
therefore conclude that the most robust diagnostic of \teff\ in RSGs
is the line-free continuum in the {\it IJHK} region, including the
$H$-hump.

\subsubsection{A connection between spectral type and evolutionary stage?}

Finally, it is worth posing the question that if the temperatures of
RSGs are all roughly constant, what is the physical interpretation of
the spectral type sequence? It has long since been assumed that the MK
classification system was associated with a decreasing temperature
scale from O to M types, with the subtypes from late-K onwards being
determined from the strengths of the TiO bands. This assumption may
well be valid for M giants and dwarfs, in which the pressure scale-heights and convective cells are much smaller with respect to the size
of the star. However, in the case of RSGs, our results from SED
fitting show that the temperatures are not strongly correlated with
the strengths of the TiO bands (and therefore spectral
type). Inspection of Fig.\ \ref{fig:HR} reveals that the $T_{\rm TiO}$
values (a proxy for spectral type) do seem to be correlated with {\it
  luminosity}, similar to the Ca~II lines used in extragalactic
studies \citep[e.g.][]{Humphreys86}. Therefore, in the M supergiant
domain, it seems the spectral types of stars are highly sensitive to
luminosity as well as temperature. 

In discussing this further, we first make the assumption that the TiO lines, which form far above the continuum-forming layers, are sensitive to the star's pressure scale-height $h$. This parameter goes as $h \sim T/g$, 
from which it is trivial to show that $h \sim L/T^3$, or if normalised to 
the size of the star, $h/R \sim L^{0.5}/T$ (assuming constant mass). Even 
using the previous temperature scale of L05, the range in temperatures of 
RSGs is quite narrow, $\sim$3600-4400K (or $\pm$10\%). By contrast, 
evolutionary models show that one expects a RSG to increase in $L$ by a 
factor of $\sim$2 as it evolves \citep[see e.g.\ Fig.\ 8 in][]{Mey-Mae00}. Therefore, as a star ascends the RSG branch, its scale-height will increase,  driven primarily by the increasing luminosity. 

From this we suggest that RSGs with the earliest spectral types may be those objects which have just arrived in the RSG phase, having evolved from the blue. The time-averaged spectral type of a RSG will then move steadily to later types as the star gradually increases in luminosity as it evolves. This is consistent with observations of coeval clusters of RSGs, where all RSGs should have approximately the same initial mass, in which a correlation is found between MK spectral type and mid-IR excess \citep{Cohen-Gaustad73}. Larger excess emission is indicative of larger amounts of circumstellar material, which suggests higher mass-loss rates (associated with higher luminosity, and hence more evolved objects), and/or a star which has spent longer in the RSG phase.

\subsubsection{Future work}

In the present study, we have highlighted the differences and similarities between the synthesised spectra of 1-D and 3-D models. Concentrating on these similarities, we have redefined the temperatures of RSGs using the 1-D models, and provided results which should be robust to 3-D effects. In the future we intend to make a more thorough analysis of this by repeating this study with a grid of 3-D models. The switch from 1-D hydrostatic to 3-D hydrodynamic models is non-trivial as it requires a dramatic increase in computational effort, nevertheless work is currently underway to construct such a grid.  

With a suite of 3-D models we will also address the effect of any departures from spherical symmetry. Should the surface flux of the star be non-uniform, e.g.\ due to hot-spots and cool-spots, the relation of Eq.\ (\ref{equ:fim}) would no longer hold. This would then cause the observer to underestimate the size of the star for a given temperature and bolometric flux. This is important with respect to comparisons with evolutionary models, which compute an effective temperature from the size and luminosity of a star assuming spherical symmetry.


\section{Summary \& conclusions} \label{sec:conc}

We have re-appraised the temperatures of Red Supergiants (RSGs) in the
Magellanic Clouds using XSHOOTER flux-calibrated spectroscopy from
400-2500nm. We simultaneously fit effective temperature \teff\ and
extinction \AV\ by fitting the spectra with a grid of 1-D spherical
MARCS model atmospheres.

When fitting the {\it VRI} spectral region, which is dominated by the
TiO absorption which has historically been used to define these stars'
temperatures, we find temperatures much lower than from fitting the
continuum SED in the {\it IJHK} region. After exploring many avenues
of evidence, we conclude that the SED temperatures are the more
reliable, for the following reasons:

\begin{itemize}
  \item Models which are tuned to match the TiO bands overpredict the
    flux longward of $\sim$900nm, dramatically so in some cases. The
    SED fits on the other hand provide good to all regions of the
    spectrum, apart from the regions where the TiO bands dominate. 
  \item The TiO temperatures $T_{\rm TiO}$ are lower than those
    measured from comparing the integrated fluxes of the data with
    those of models (the {\it flux integration method, FIM}), which is
    the least model-dependent method of determining temperature. Even
    if we assume minimal reddening, the FIM temperatures $T_{\rm FIM}$
    are higher than those from the TiO fits. The FIM temperatures do
    agree with the SED fits if we assume a modest amount of
    extinction, an entirely reasonable assumption.
  \item The implied total line-of-sight extinctions from the TiO fits
    are in most cases {\it lower} than those of nearby stars, even
    though RSGs are expected to have higher extinction due to
    circumstellar material. By contrast, the SED fits imply extinctions
    which are in agreement with or in excess of that of the nearby
    stars. 
  \item The extinction from the TiO fits seems to be completely
    uncorrelated with the strengths of the diffuse interstellar band at
    862nm, whereas the extinction from the SED fits appears to show a
    degree of correlation. 
\end{itemize}

To reconcile these two measurements of temperature, we considered the
effects of overall metallicity, CNO mixture, and dynamical effects
(convection and granulation). Of these we conclude that the latter is
most likely cause. Recent studies have shown that 3-D hydrodynamical
models of RSGs have a significantly different radial temperature
profile in the upper layers of the atmosphere where the TiO lines
form. The temperature structure at deeper layers, where the continuum
forms, is largely unchanged. For this reason we consider the {\it
  IJHK} continuum region and the \Hminus\ opacity minimum at
1.6\um\ to be robust temperature indicators. 

Adopting the SED temperatures, we find a number of interesting
effects. Firstly, the temperatures of all RSGs in both Magellanic
Clouds are roughly uniform, 4150$\pm$150\,K. This brings the observed
Hayashi limit for massive stars into agreement with those of Geneva
models, and somewhat warmer than those of Bonn and Cambridge
models. This also means that the strengths of the TiO bands, which
define the spectral types of M supergiants, may also strongly depend on luminosity. From this we suggest that the spectral type of a RSG may be indicative of its evolutionary state, with more evolved RSGs having later spectral types.

Secondly, the circumstellar extinction of RSGs is weakly correlated
with luminosity, consistent with recent findings that more luminous
stars have stronger winds. We also see a separation between LMC and
SMC stars, implying a metallicity dependence on wind strength.

Thirdly, as there is very little variation in temperature of the RSGs,
their bolometric corrections are all approximately the same. This
means that, if we can account for extinction, we can determine a RSG's
bolometric luminosity to a high degree of precision from a single
photometric point. This has applications for the study of supernova
progenitors, where pre-explosion imaging is often limited to one or two
photometric bands. We have shown that the circumstellar extinction is
very low in the near-infrared, so if the foreground extinction is
known (e.g. from neighbouring stars) the luminosity can be measured to
a precision of $\pm$0.1dex from only one photometric band. In the optical
({\it VRI}) region, circumstellar extinction becomes more significant,
as does the effect of TiO absorption on the broad-band flux. However,
since both these effects are luminosity dependent, we have been able
to provide a calibration which recovers the luminosity from optical
photometry to a precision of $\pm$0.2dex.

\acknowledgments Acknowledgments: BD is supported by a fellowship from
the Royal Astronomical Society. The data used in this paper is from
ESO programme 088.B-0014(A). This work was supported in part by the National
Science Foundation under grant AST-1108906 to RPK. Moreover, RPK
acknowledges support by the Alexander-von-Humboldt Foundation and the
hospitality of the Max-Planck-Institute for Astrophysics in Garching
and the University Observatory Munich, where part of this work was
carried out. For our analysis we used the software package IDL, the The IDL Astronomy User's Library at GSFC, and the Coyote graphics library. Finally, we thank Sylvia Ekstrom, John Eldridge, Roberta
Humphreys, Norbert Langer, Emily Levesque, Phil Massey, Georges
Meynet, and Chris Tout for useful discussions. 

\section{Appendix}

\begin{figure*}
  \centering
  \includegraphics[width=17cm]{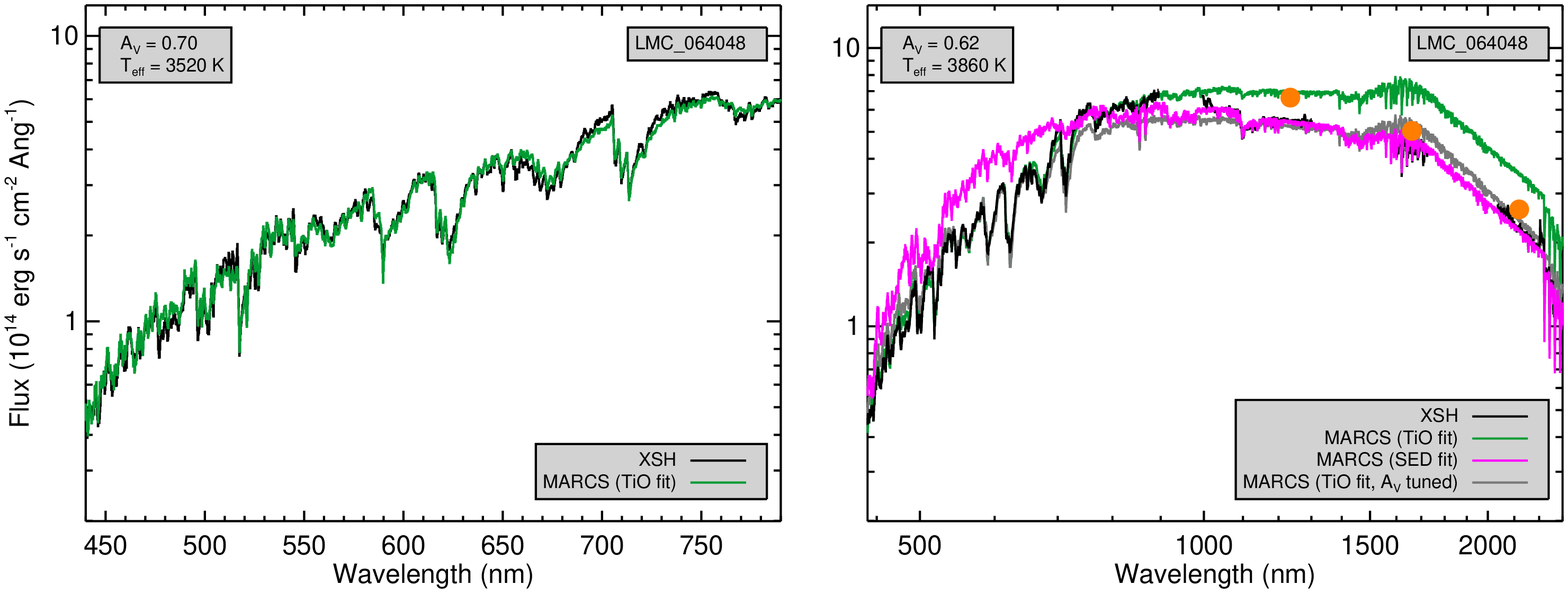}
  \includegraphics[width=17cm]{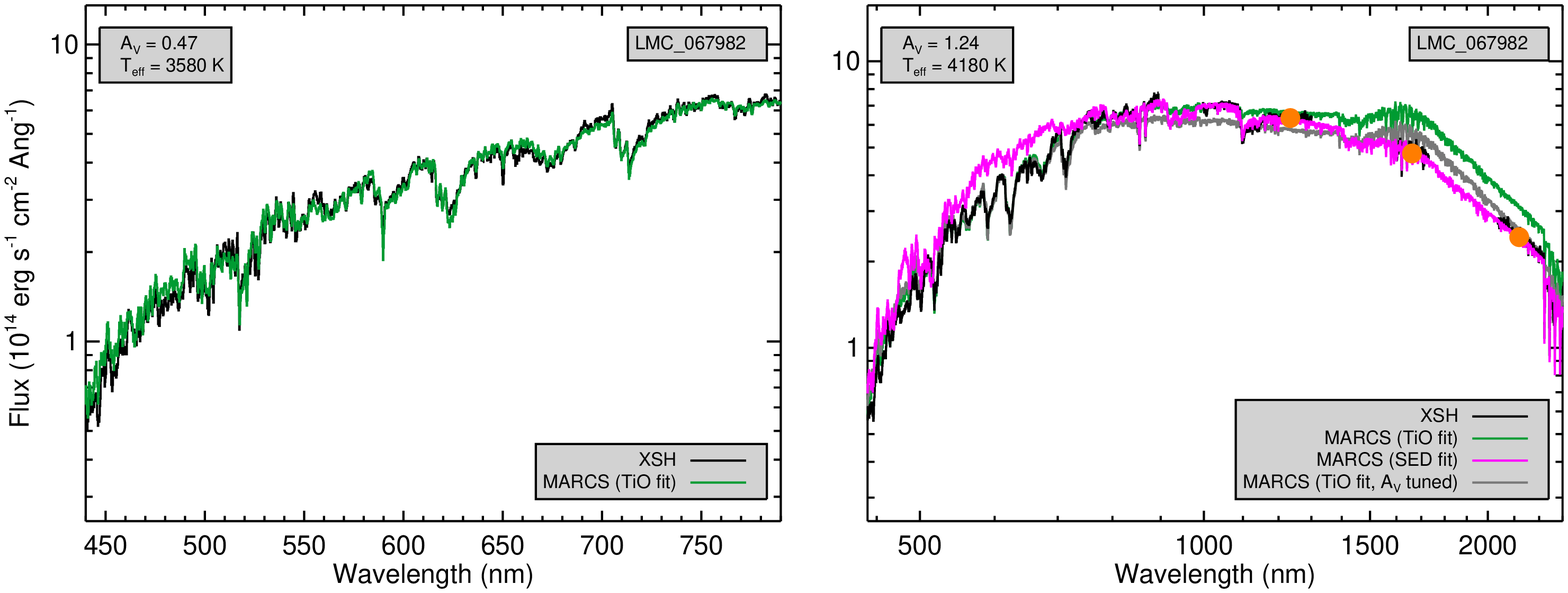}
  \includegraphics[width=17cm]{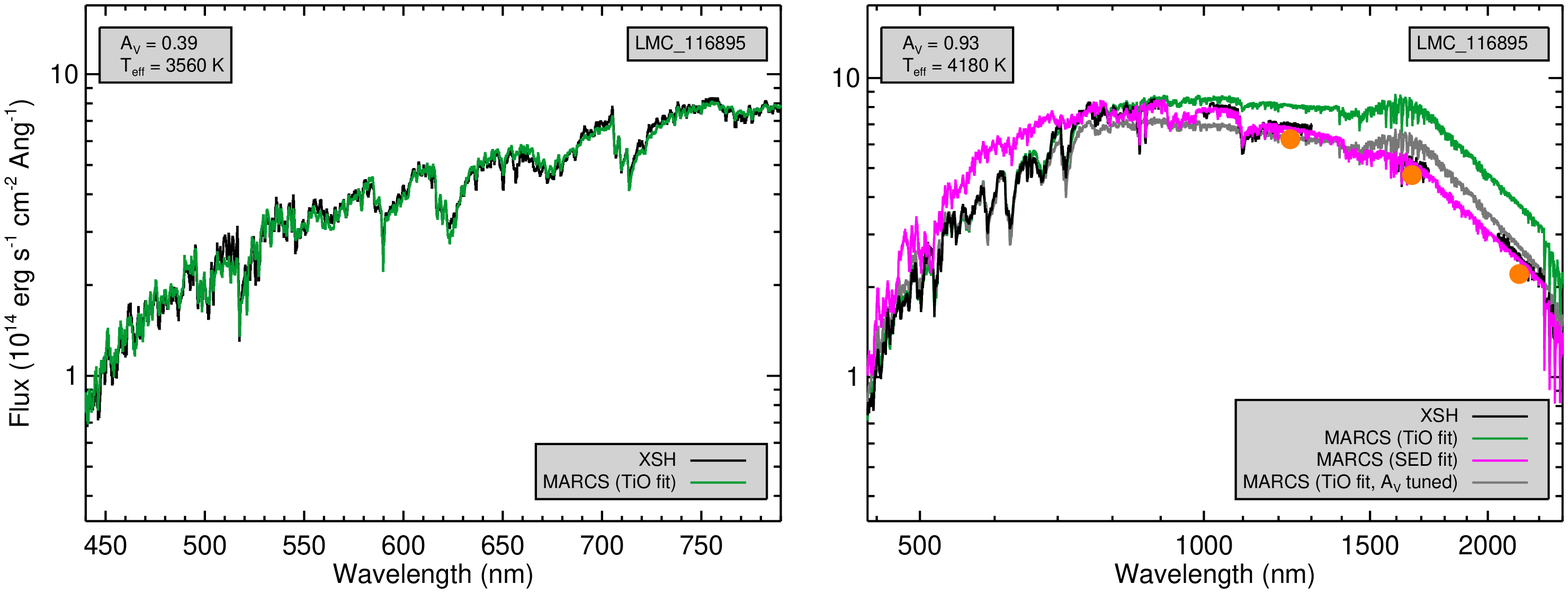}
  \caption{MARCS model fits to the XSHOOTER data. The left panels show
    the observations (black) and the best fitting model (green), along
    with the fit parameters in the top right. The right panels show
    the same data over a broader wavelength range (black), the best
    SED fit (magenta), the TiO bands fit from the left panel (green),
    and the best fitting model with the \Teff\ from the TiO bands fit
    but which has had the extinction tuned in order to give the best
    fit to the whole SED (grey). Overplotted in orange are the 2MASS
    photometry for each star. }
  \addtocounter{figure}{-1}
  \label{fig:example}
\end{figure*}
\begin{figure*}
  \centering
  \includegraphics[width=17cm]{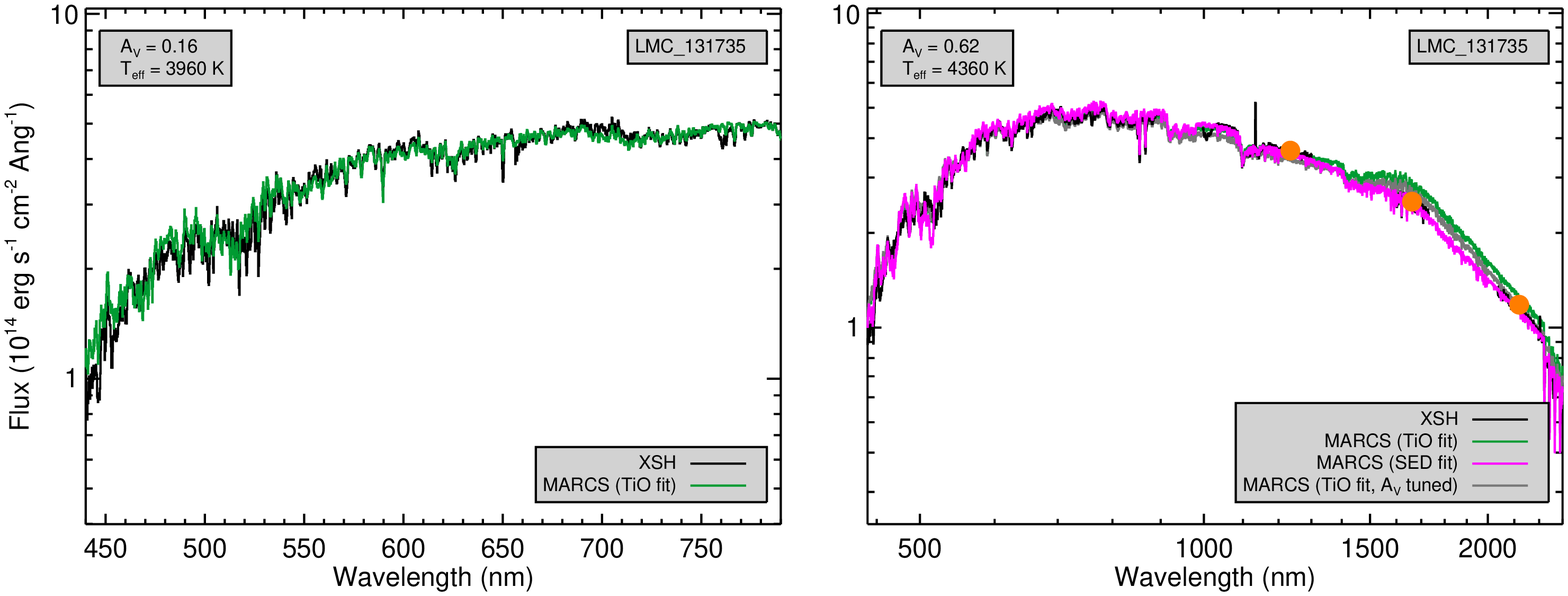}
  \includegraphics[width=17cm]{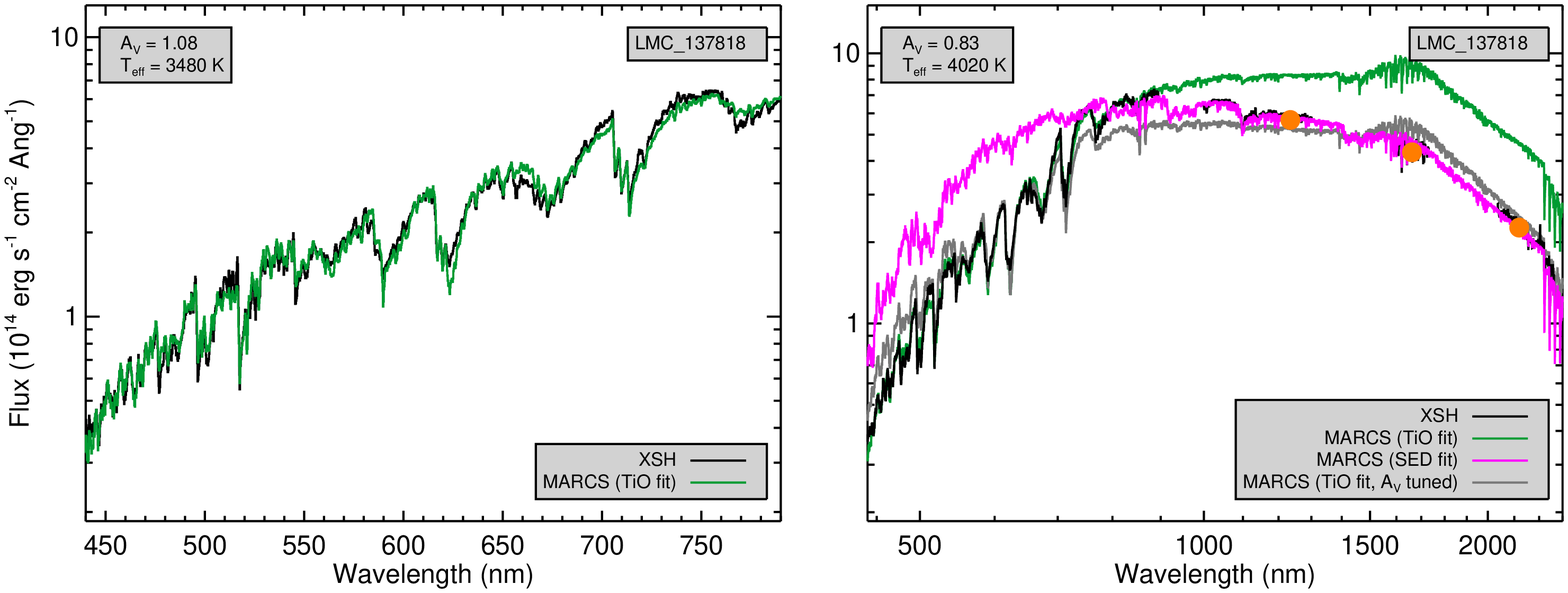}
  \includegraphics[width=17cm]{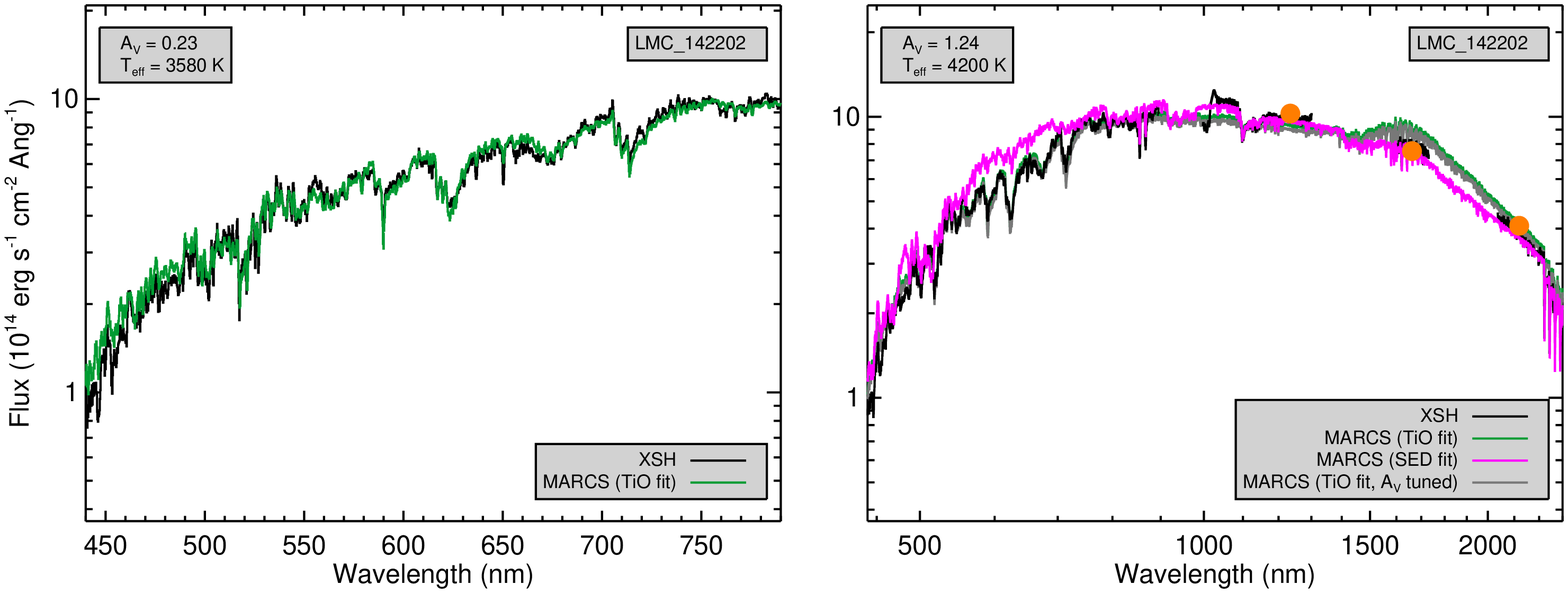}
  \caption{Continued.  }
  \addtocounter{figure}{-1}
\end{figure*}
\begin{figure*}
  \centering
  \includegraphics[width=17cm]{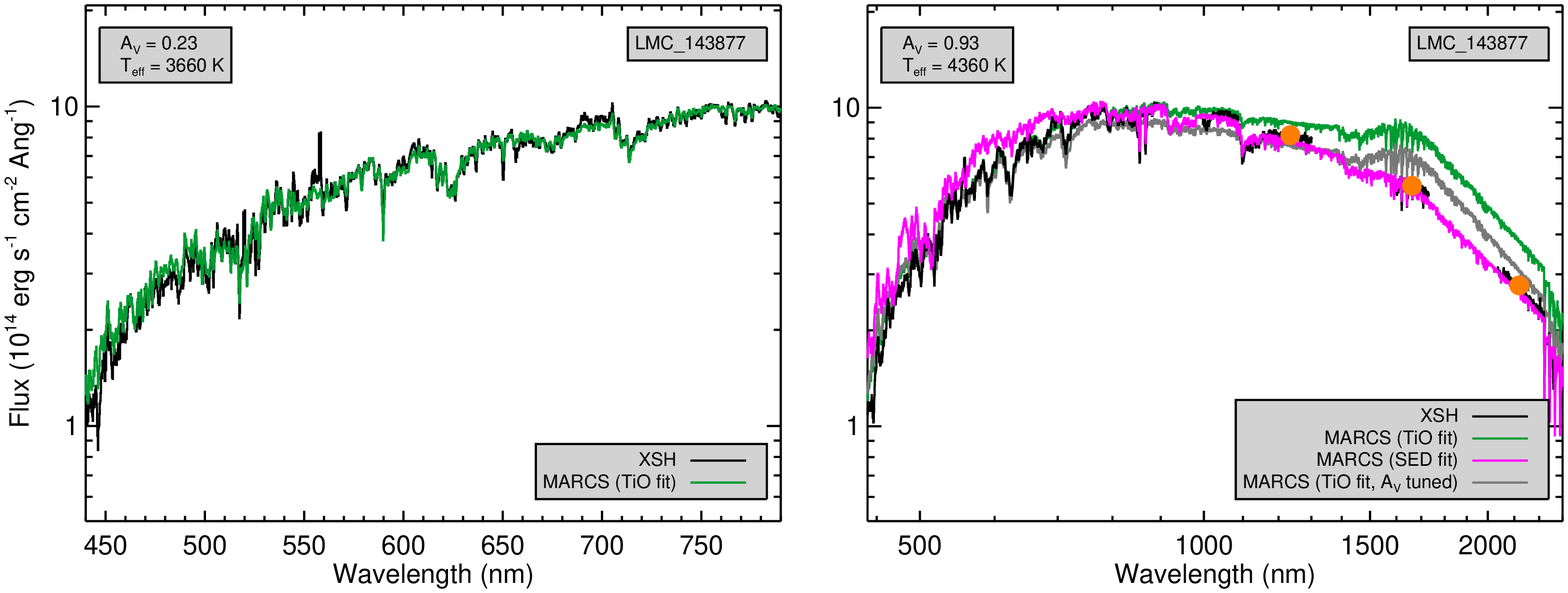}
  \includegraphics[width=17cm]{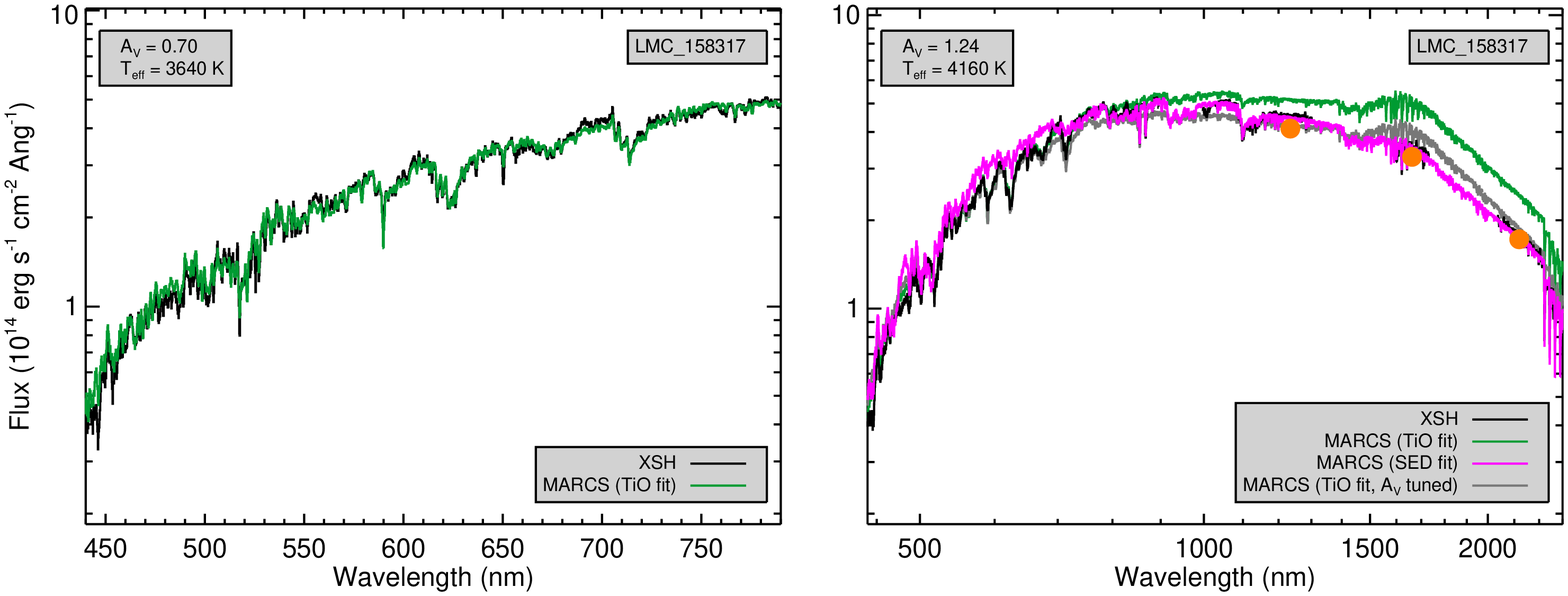}
  \includegraphics[width=17cm]{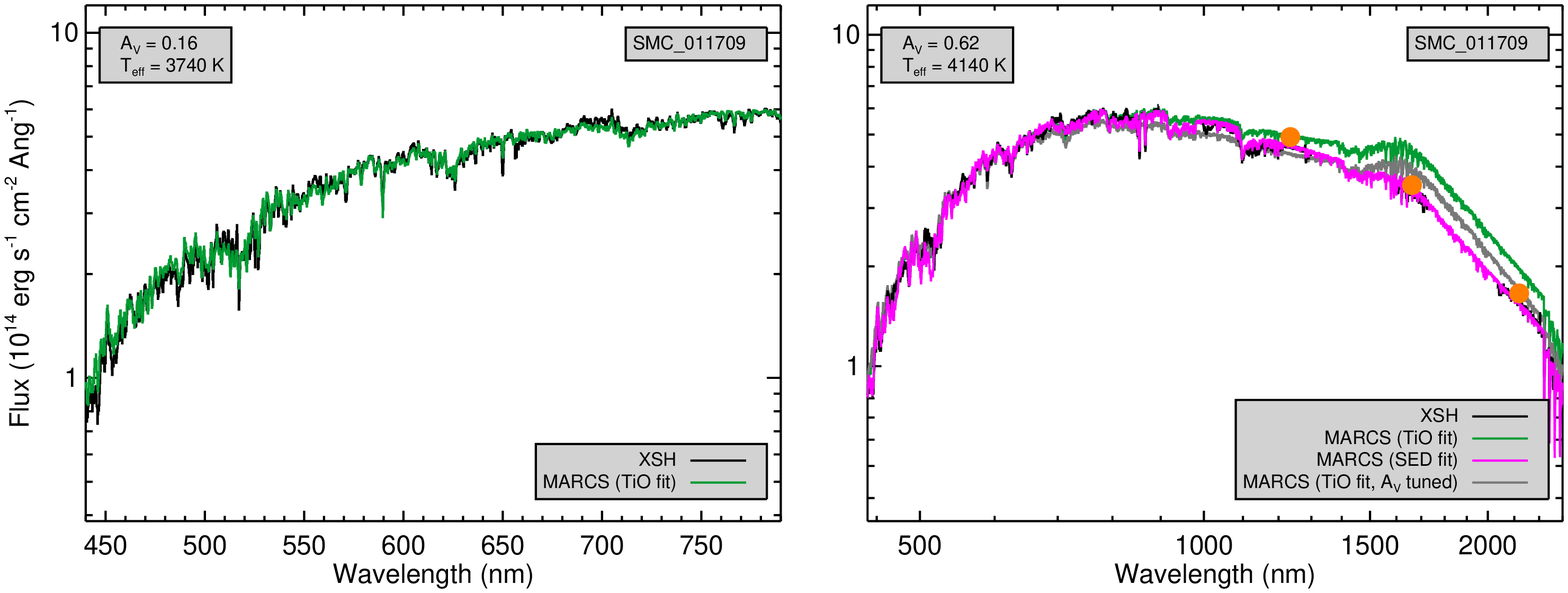}
  \caption{Continued.  }
  \addtocounter{figure}{-1}
\end{figure*}
\begin{figure*}
  \centering
  \includegraphics[width=17cm]{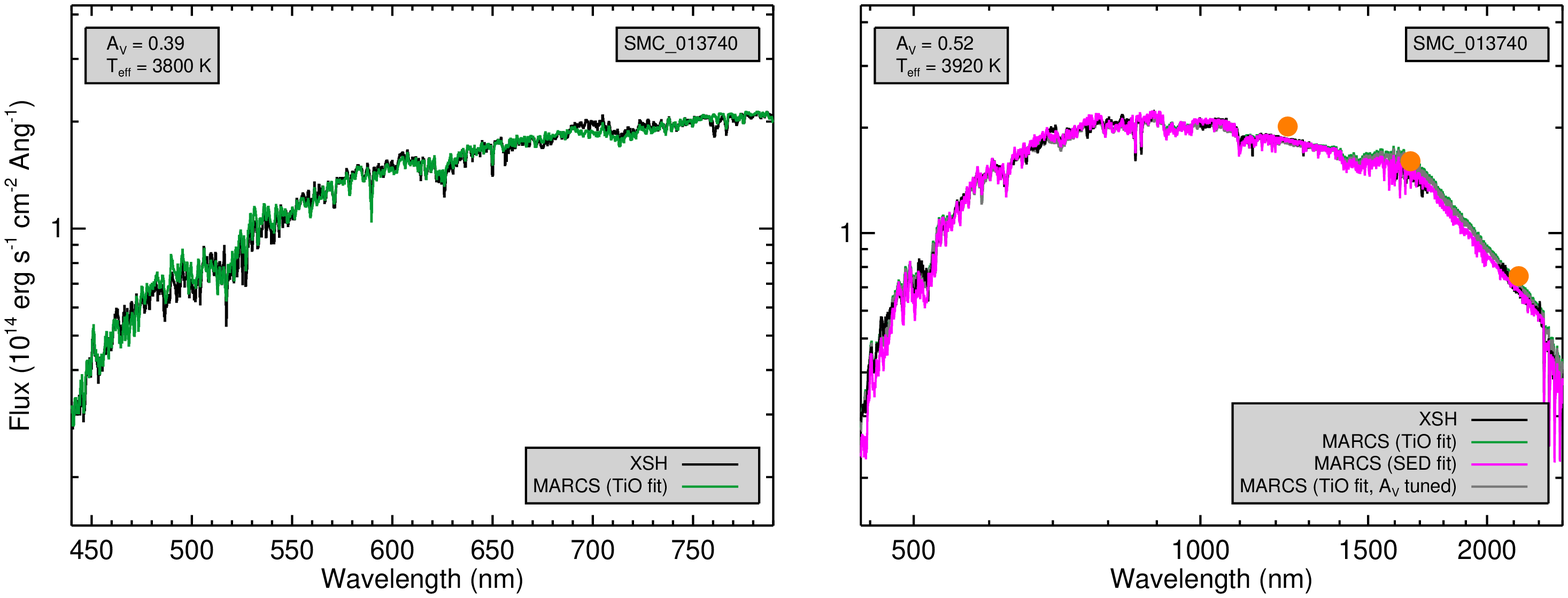}
  \includegraphics[width=17cm]{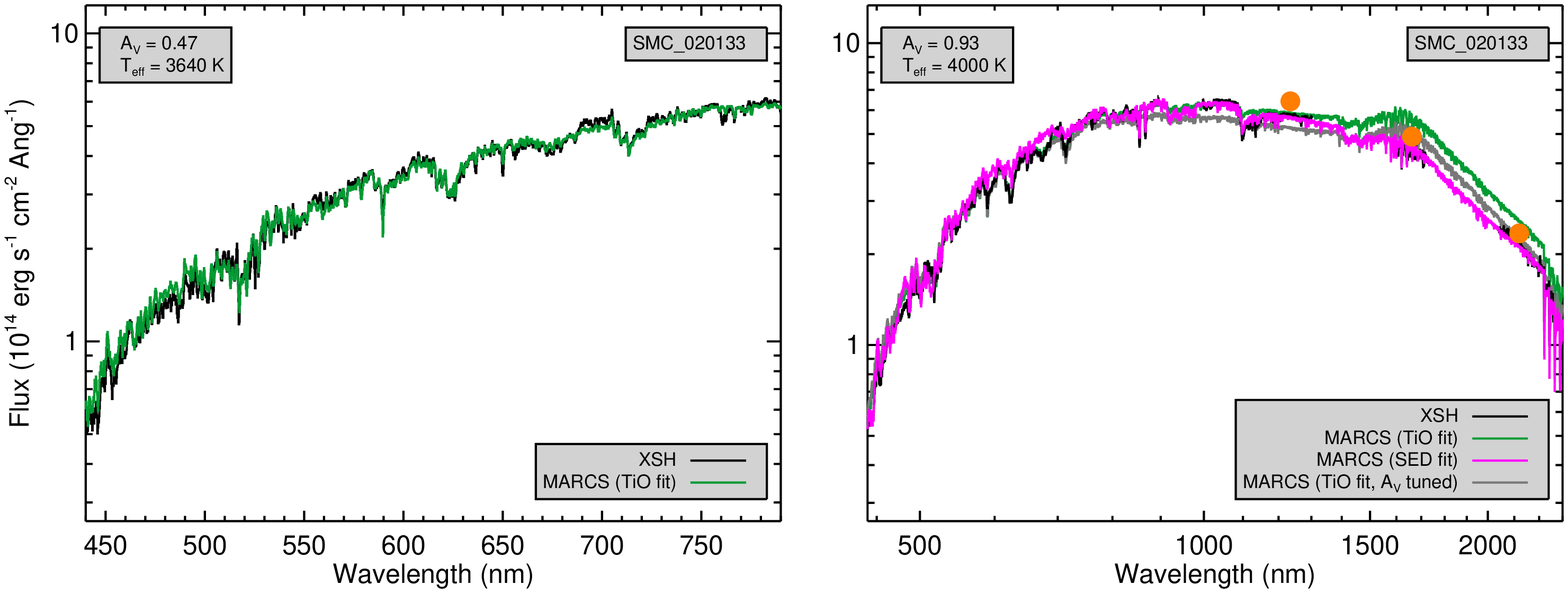}
  \includegraphics[width=17cm]{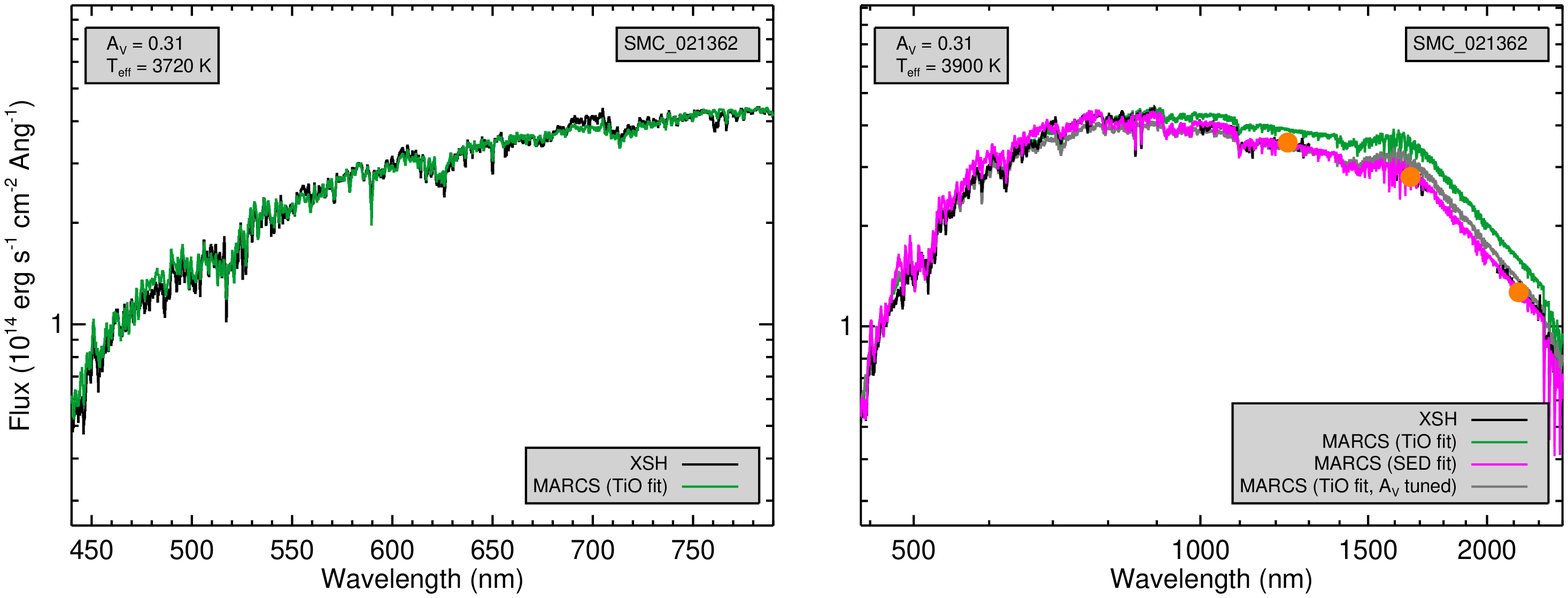}
  \caption{Continued.  }
  \addtocounter{figure}{-1}
\end{figure*}
\begin{figure*}
  \centering
  \includegraphics[width=17cm]{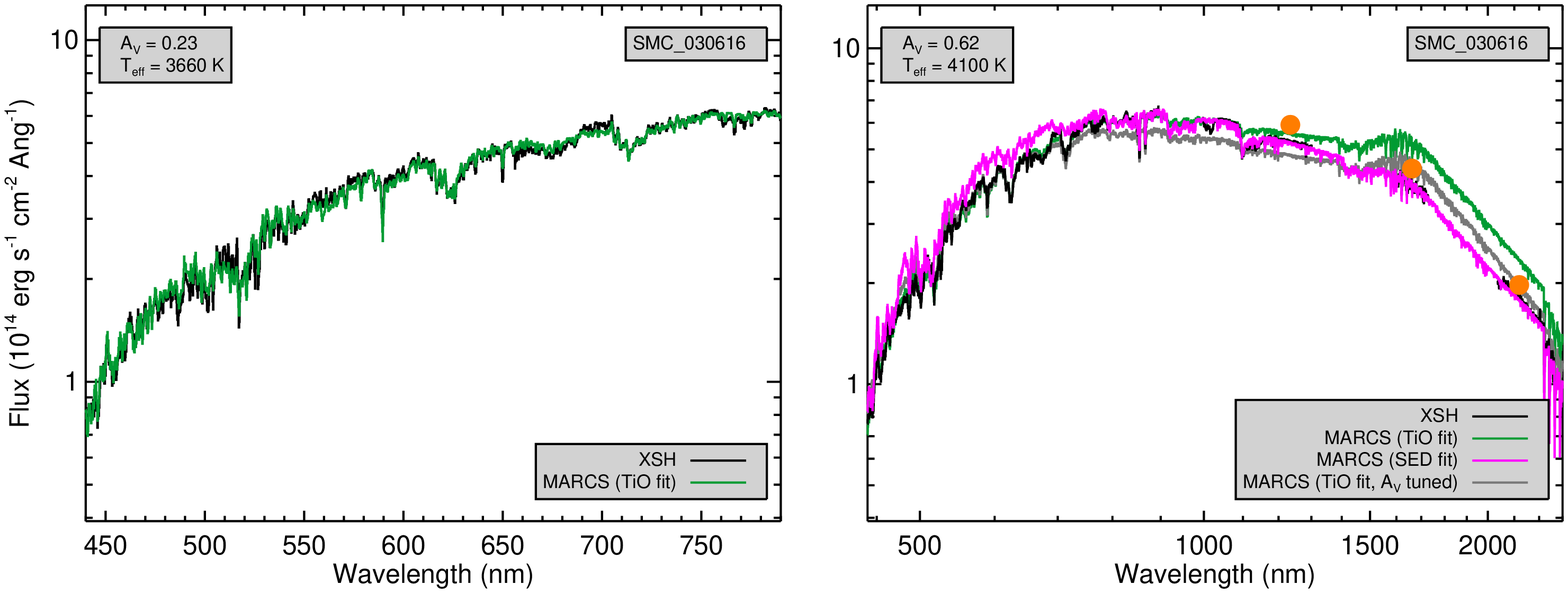}
  \includegraphics[width=17cm]{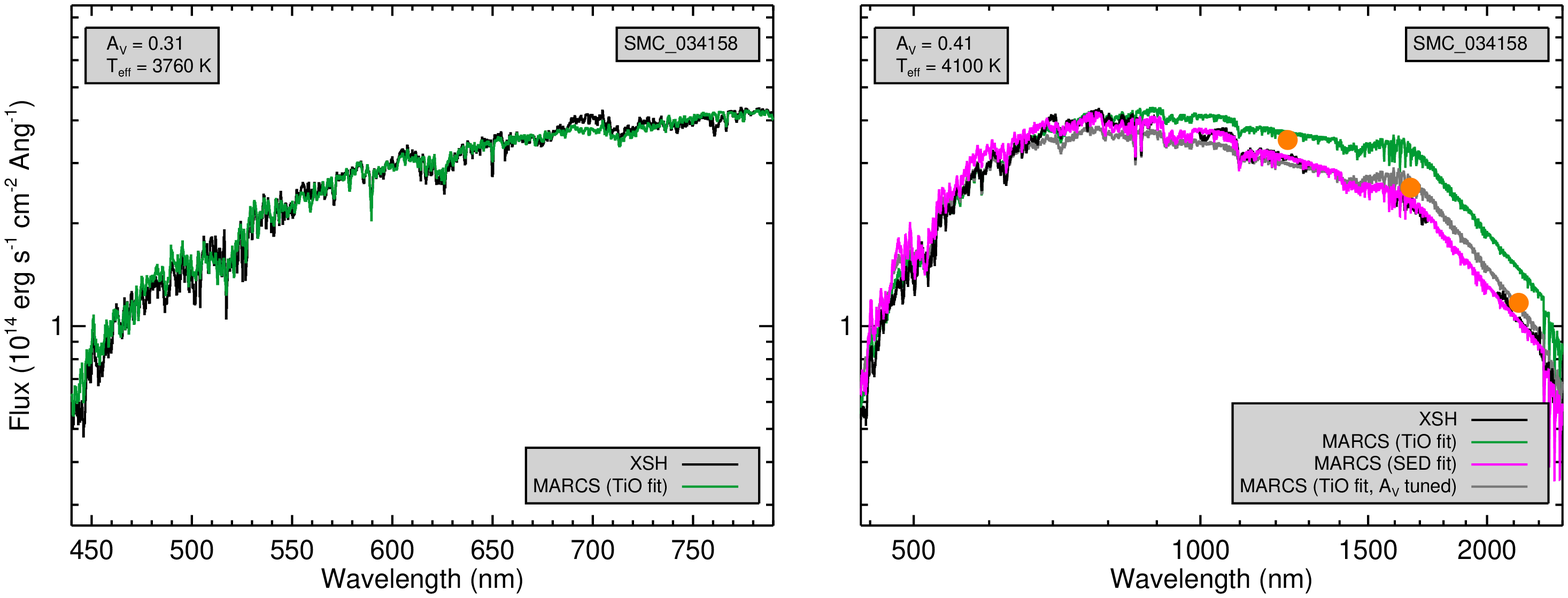}
  \includegraphics[width=17cm]{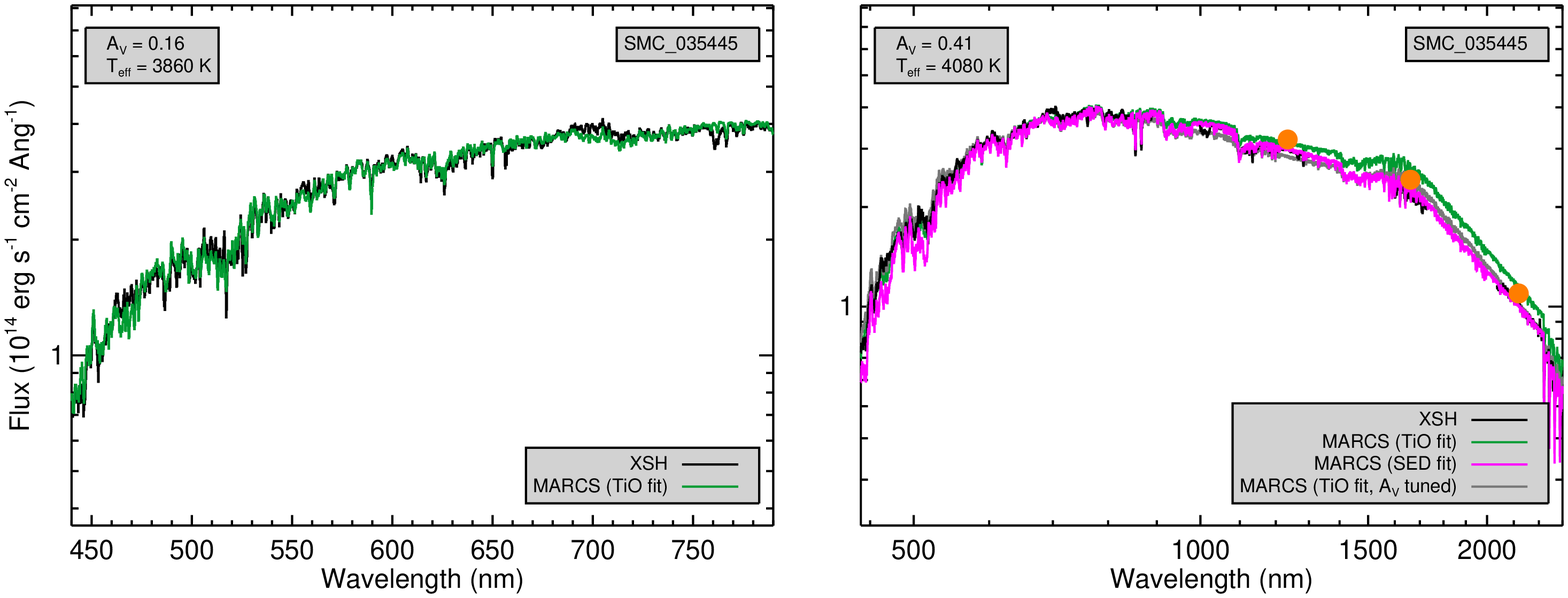}
  \caption{Continued.  }
  \addtocounter{figure}{-1}
\end{figure*}
\begin{figure*}
  \centering
  \includegraphics[width=17cm]{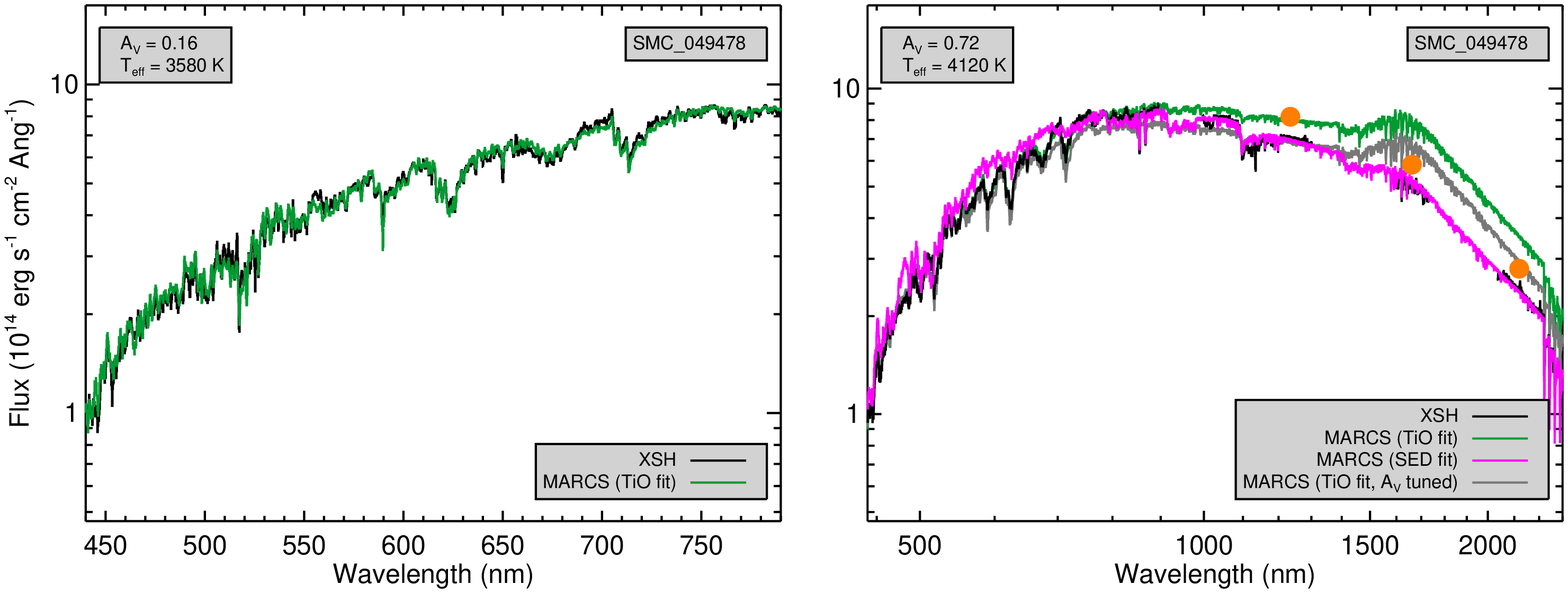}
  \includegraphics[width=17cm]{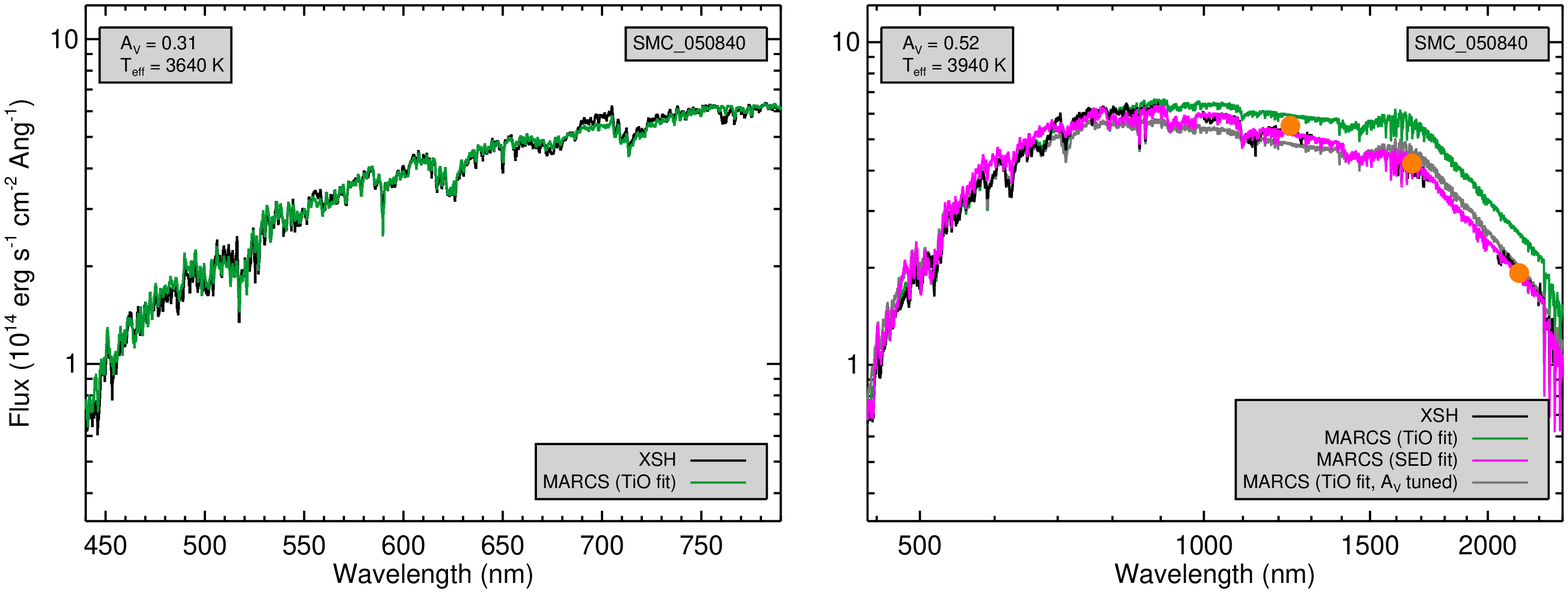}
  \includegraphics[width=17cm]{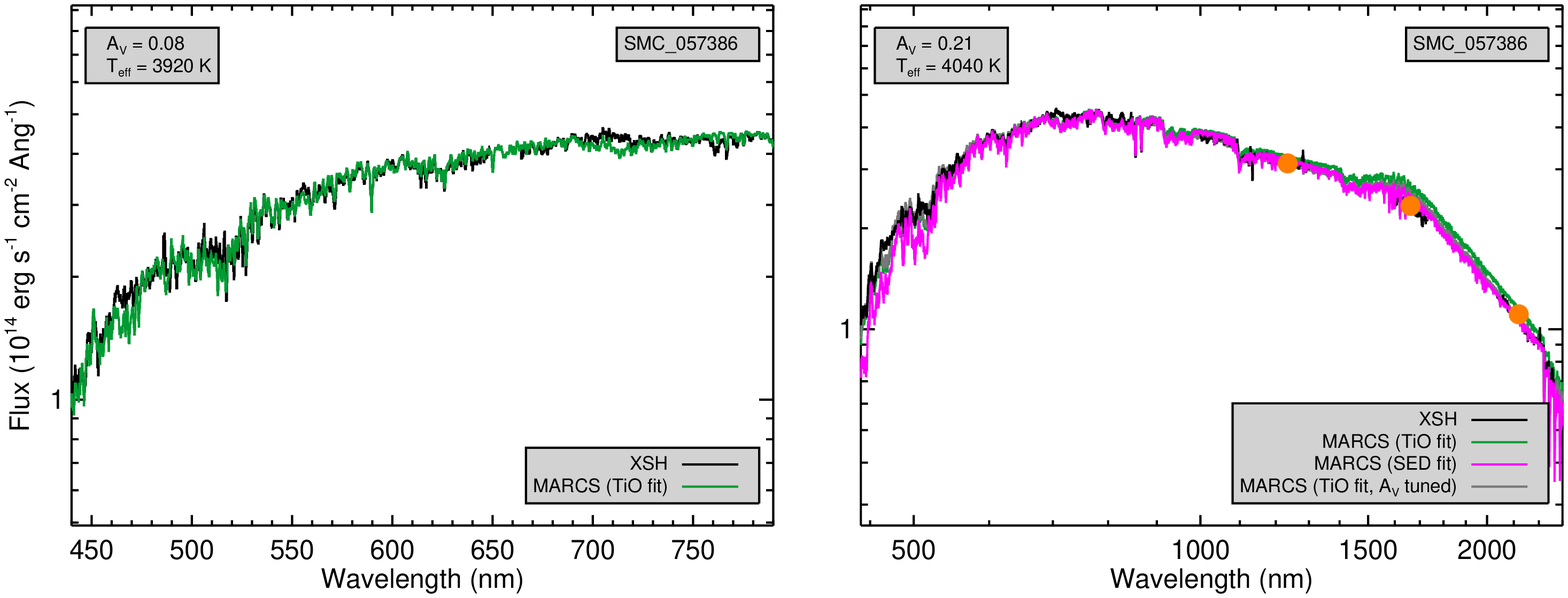}
  \caption{Continued.  }
  \addtocounter{figure}{-1}
\end{figure*}

\bibliographystyle{/fat/Data/bibtex/astronat/apj/apj.bst}
\bibliography{/fat/Data/bibtex/biblio}

\end{document}

%% file: journals.tex
\def\apjl{ApJ}%
\def\apjs{ApJS}%
\def\ao{Appl.~Opt.}%
\def\aap{A\&A}%